\documentclass[sigconf,nonacm,balance=false]{acmart}

\usepackage[all]{nowidow}
\usepackage{siunitx}
\usepackage{multirow}
\usepackage{amsmath}
\usepackage{enumitem}
\usepackage{tcolorbox}
\usepackage[linesnumbered,ruled,vlined]{algorithm2e}
\usepackage{caption}
\usepackage{subcaption}
\usepackage{colortbl}
\usepackage{diagbox, eqparbox, hhline}
\usepackage{hyperref}
\usepackage{booktabs}

\usepackage{booktabs}
\usepackage{tabularx}
\usepackage{multirow}
\usepackage{makecell}
\usepackage{listings}
\usepackage{xurl}

\SetCommentSty{mycommfont}

\usepackage{wasysym}
\newcommand{\fullcirc}{\CIRCLE}
\newcommand{\emptycirc}{\Circle}

\newcommand{\cmark}{\checkmark}
\newcommand{\xmark}{}

\usepackage{xspace}
\makeatletter
\DeclareRobustCommand\onedot{\futurelet\@let@token\@onedot}
\def\@onedot{\ifx\@let@token.\else.\null\fi\xspace}

\newcommand{\eg}{e.g., }
\newcommand{\ie}{i.e., }
\newcommand{\etc}{etc.}

\def\etc{etc\onedot}

\makeatother

\newlist{rqlist}{enumerate}{3}
\setlist[rqlist]{label=\textbf{RQ\arabic*}.,before=\raggedright,leftmargin=40pt,ref=RQ\arabic*}

\newcounter{result}
\newcommand{\result}{\refstepcounter{result}\textbf{Result~\arabic{result}}}

\newcommand{\shortsection}[2][.]{\noindent\textbf{#2#1}}

\lstdefinelanguage{CodeQL}{
  keywords={import, from, where, select, class, extends, predicate, exists},
  sensitive=true,
  comment=[l]{//},
  morecomment=[s]{/*}{*/},
  morestring=[b]"
}

\newcommand{\NumCVEs}{3\;993\xspace}
\newcommand{\NumDetectedCVEs}{171\xspace}
\newcommand{\NumDetectedCVEsInTime}{83\xspace}
\newcommand{\NumCVEsRan}{3,851\xspace}
\newcommand{\PercentDetectedCVEsInTime}{48\%\xspace}

\newif{\ifanonymous}

\begin{document}

\title[]{Longitudinal Analyses of SAST Tools: A CodeQL Case Study}

\ccsdesc[500]{Security and privacy~Software and application security}
\ccsdesc[500]{Software and its engineering~Software evolution}

\keywords{CodeQL, Static Analysis Tools, Vulnerability Detection}

\author{Jean-Charles Noirot Ferrand}
\orcid{0009-0009-9650-4011}
\affiliation{%
  \institution{University of Wisconsin-Madison}
  \city{Madison}
  \state{Wisconsin}
  \country{USA}}
\email{jcnf@cs.wisc.edu}

\author{Kyle Domico}
\orcid{0009-0002-7853-4910}
\affiliation{%
  \institution{University of Wisconsin-Madison}
  \city{Madison}
  \state{Wisconsin}
  \country{USA}}
\email{domico@cs.wisc.edu}

\author{Yohan Beugin}
\orcid{0000-0003-0991-7926}
\affiliation{%
  \institution{University of Wisconsin-Madison}
  \city{Madison}
  \state{Wisconsin}
  \country{USA}}
\email{ybeugin@cs.wisc.edu}

\author{Patrick McDaniel}
\orcid{0000-0003-2091-7484}
\affiliation{%
  \institution{University of Wisconsin-Madison}
  \city{Madison}
  \state{Wisconsin}
  \country{USA}}
\email{mcdaniel@cs.wisc.edu}

\renewcommand{\shortauthors}{Noirot Ferrand et al.}

\begin{abstract}
Open-source software (OSS) pipelines rely on automated static analysis tools to prevent the introduction of vulnerabilities in code.
However, there is limited understanding of the efficacy of these tools across the OSS ecosystem over time.
In this paper, we introduce a novel method to evaluate static application security testing (SAST) tools through longitudinal measurements and perform the largest academic study of CodeQL---the most prevalent static analysis tool from GitHub---on OSS codebases.
We apply our apparatus on 114 versions of CodeQL over time on 3\;993 CVEs from 1\;622 repositories to measure key properties of the tool, culminating in more than 20 billion lines of code analyzed. First, we measure its effectiveness, \ie its ability to detect vulnerabilities before they are fixed. Then, we determine whether these detections were actionable through two measures of the distance between findings and vulnerability location either over the entire codebase or within the vulnerable file. Finally, we study the stability of CodeQL by examining how vulnerability detections hold across versions and the evolution of CodeQL on the accuracy-precision trade-off.
We find that CodeQL identifies a total of 171 CVEs, and that for 83 of them, a CodeQL version prior to the fix could detect it. Such detections are in general actionable if findings are triaged across files, as for 50\% of the 171 detections, more than 50\% of findings in the vulnerable file are located in the vulnerable location. Finally, we show that CVE detections are not monotonic across versions as 21 CVEs were no longer detected following a version change and 17 that were never redetected.
Our study shows that using SAST tools is a matter of best practice as they prevent numerous vulnerabilities from being introduced, but that developers should be aware of changes that may leave blind spots in detections upon updates of the tool.
\end{abstract}

\maketitle{}

\section{Introduction}\label{introduction}

Open-source software (OSS), as part of an ever-growing supply chain, is prone to attacks~\cite{ohmBackstabbersKnifeCollection2020,ladisaSoKTaxonomyAttacks2023}. Events such as the XZ Utils backdoor~\cite{linsCriticalPathImplant2024} or Log4shell~\cite{eversonLog4shellRedefiningWeb2022} illustrate the existence and importance of security incidents within this supply chain. To eliminate as many attack vectors as possible, software projects undergo automated static analysis, relying to this end on static application security testing (SAST) tools. In particular, CodeQL~\cite{coolAnnouncingGitHubSecurity2019}, one of GitHub Advanced Security products, is the most prevalent SAST tool. Given a source code to analyze, CodeQL first builds a database of facts about the code (capturing the program's semantics), runs a predetermined set of queries looking for specific weaknesses on this database, and outputs a set of \textit{alerts} raised by queries at different locations in the code. Introduced in 2019, its adoption has been steadily increasing as GitHub reported storing CodeQL databases for more than 200\;000 repositories in 2022~\cite{CodeQLVSCode2022}.

SAST tools such as CodeQL have undergone numerous quantitative (performance measurement)~\cite{lippEmpiricalStudyEffectiveness2022,charoenwetEmpiricalStudyStatic2024,shenFinding709Defects2025} and qualitative (user studies)~\cite{amiFalseNegativeThat2024,bennettDevelopersUseStatic2024,liUnderstandingIndustryPerspectives2025} evaluations. However, these studies are limited in scope: they evaluate a specific version of the tool(s) on a restrained set of languages and common weakness enumerations (CWEs). Since SAST tools continuously evolve, the results gathered from evaluating one version may not be the same in the next version, and no study has tracked how tool performance changes across versions. This limitation highlights a fundamental flaw in existing evaluations, as developers generally rely on the \textit{latest} version of SAST tools~\cite{bennettDevelopersUseStatic2024}, thus exposing the software to any breaking changes introduced in the release of a new version. Therefore, best practices on using SAST tools have largely been drawn from limited evidence (\ie studies on specific version, language, or CWE) which hinders generalization to the entire OSS ecosystem.

In this paper, we posit that best practices on SAST can only be derived from a rigorous evaluation on the numerous axes that influence results: OSS ecosystems (languages), vulnerability types (CWEs), and most importantly the development of the SAST tool itself (versions). We propose a new framework to determine the effectiveness of a SAST tool, the actionability of its generated alerts, and their stability across versions.
We instantiate this method to measure the practicality of CodeQL for vulnerability detection. We apply different releases (versions) of CodeQL on known CVEs and evaluate if they detect them. Using the release dates of CodeQL and the fix commit date of the CVEs, we characterize the lead time of CodeQL over a CVE: the duration between its fix commit date and the date of the first CodeQL release that detects it. Then, we characterize whether such detections were actionable at the time by measuring the location of alerts with respect to the vulnerability location at two levels. Finally, we measure if alerts that detect CVEs are stable over versions (\ie consistently detects the CVEs in subsequent versions) and how the utility-cost trade-off of the tool evolved over time because of modifications of the tool.

We perform the largest known academic evaluation of a SAST to the best of our knowledge by applying our approach on CodeQL onto a subset of CVEfixes~\cite{bhandariCVEfixesAutomatedCollection2021}---a dataset of CVEs and the commits that fixed them. Our analysis spans 3\;993 CVEs of 1\;622 unique repositories for 6 languages and 114 versions of CodeQL (between v2.7.0 and v2.25.2), for a total of more than 100\;000 executions of the tool, more than 20B lines of code analyzed per version and close to 3M alerts generated in total. A CVE is considered detected prior to the fix if a subset of alerts matches one of the fix commit edits and no longer appears after them. Then, for detected CVEs, we measure the location of the generated alerts with respect to the vulnerable location at the project-level (with respect to the vulnerable files) and the file-level (with respect to the vulnerable lines).

We find that out of the \NumCVEs CVEs we analyzed, CodeQL identifies \NumDetectedCVEs of them. Importantly, \PercentDetectedCVEsInTime (\NumDetectedCVEsInTime) of these CVEs could have been detected prior to their fix. The generated alerts for the detected CVEs are largely sparse at the project level (\ie not concentrated in the vulnerable files): for 50\% of vulnerability detected, only 10\% of alerts are located in the vulnerable file(s). However, if a developer would adopt a triage strategy that isolates the most likely vulnerable file, the detections would be more actionable, as the majority of alerts in vulnerable files overlap with the vulnerability's location. Finally, we found that vulnerabilities are not guaranteed to be detected in subsequent versions (\ie detections are non-monotonic) as 21 detected CVEs had at least one version which could not detect them while the previous version could. Among those 21 CVEs, 17 were permanently dropped. In particular, 2 of those permanent drops are linked to a product strategy (and not simple precision changes): GitHub removed the \texttt{hardcoded-credentials} from CodeQL when GitHub Secret Protection appeared as a separate product, which left developers and projects unaware of this change exposed to such vulnerabilities. Therefore, relying on such tools warrants for awareness of their changes which can negatively impact their effectiveness and the security coverage they provide.

Our contributions are as follows:
\begin{itemize}
    \item We introduce a new framework to characterize the performance of a SAST tool through longitudinal measurements.
    \item We perform the largest academic study on a SAST tool, resulting in more than 20B LoC analyzed per CodeQL version.
    \item We show that using CodeQL would have detected \NumDetectedCVEsInTime CVEs before the time of the fix.
\end{itemize}
\section{Background}\label{section:background}\label{section:background-codeql}

\subsection{CodeQL}
CodeQL---the successor to LGTM after the acquisition of Semmle by GitHub in 2019---is an open-source static analysis tool that is part of the GitHub Advanced Security suite. It is used by developers as a way to secure their code~\cite{bennettDevelopersUseStatic2024} and by academics for evaluation~\cite{charoenwetEmpiricalStudyStatic2024} or focused studies~\cite{shenFinding709Defects2025,tiwariItsTimeEmpirical2025} as it enables the creation of specific queries that leverage the tool's code extraction capabilities. It is powered by the .QL language~\cite{deMoor2008} introduced by Semmle in 2007, an object-oriented query language designed to retrieve information from relational databases.

\shortsection{Versions} As any software, CodeQL is regularly updated to improve the ecosystem support (new languages and their versions), the security coverage (new queries), the overall tool (new features or bug fixes), or adjust the precision of the tool. This results in a new \textit{version}: a snapshot of the current command-line interface (CLI) and the companion repository containing the libraries and queries. Each version is assigned a version number following the pattern \texttt{vX.Y.Z}\footnote{In practice, this start at v2.4.0 because CodeQL was open-sourced after the acquisition.} where \texttt{X}, \texttt{Y}, and \texttt{Y} represents the major, minor, and patch version numbers, respectively. On average, patches are released every two weeks and minor versions are releases every 3 months.

\shortsection{Usage in practice} While it is possible to use CodeQL manually through the CLI (\eg security researchers who craft specific queries), most projects use it as part of their CI/CD pipelines or using the dedicated VS Code extension. GitHub exposes the two GitHub Actions \texttt{github/codeql-action/init} to set up the environment for CodeQL and \texttt{github/codeql-action/analyze} to apply the analysis. Those actions are regularly updated to use the latest available release by default.

\begin{figure}
    \centering
    \includegraphics[width=\linewidth]{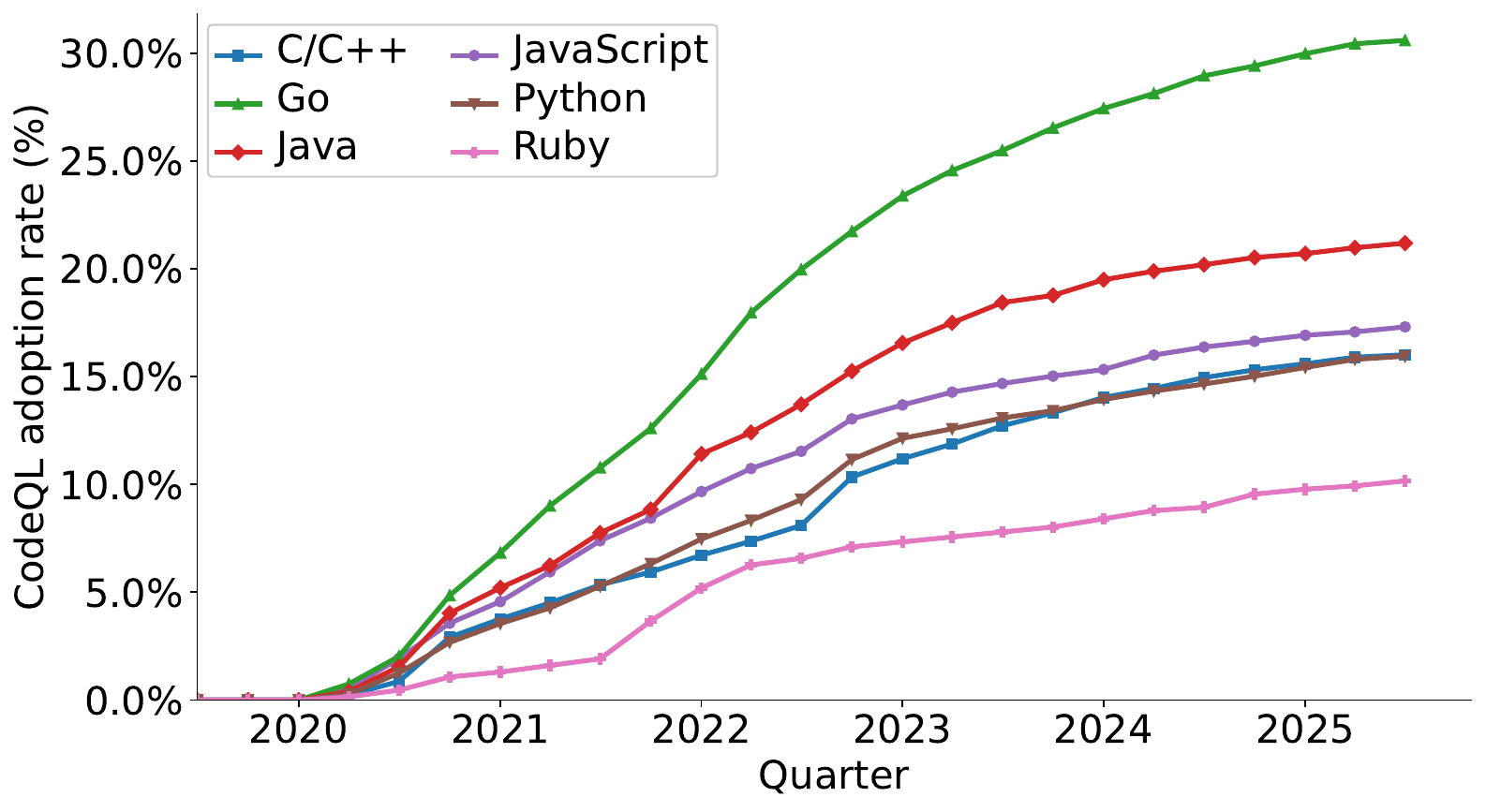}
    \caption{Adoption of CodeQL across languages}
    \Description[CodeQL adoption rates (0--31\%) for six languages from 2019 to 2025, with Go dominating.]{All six languages start near 0\% in late 2019 (introduction of CodeQL) and grow monotonically with the steepest gains between 2020 and 2022. Go far outpaces the rest, reaching approximately 31 by late 2025. Java follows at approximately 21\%, then JavaScript at approximately 17.5\%. C/C++ and Python converge at approximately 16\%. Ruby shows the smallest adoption at approximately 10.5\%.}
    \label{fig:codeql-adoption}
\end{figure}

\shortsection{Adoption} \autoref{fig:codeql-adoption} shows the proportion of GitHub repositories with a CodeQL workflow, measured using the dataset from Cardoen et al.~\cite{cardoenDatasetGitHubActions2024} containing GitHub Actions workflows histories for close to 50\;000 repositories. It is clear from the graph that there was no sudden adoption. Instead, both the number and the proportion of repositories using CodeQL have been steadily increasing (albeit at a slower velocity than the first years). Further, with current adoption rates between 10\% and 30\% across languages, CodeQL establishes itself as one of the most popular, if not the most popular SAST tool.

\subsection{CodeQL Analyses}
\begin{figure}[h]
    \centering
    \includegraphics[width=\linewidth]{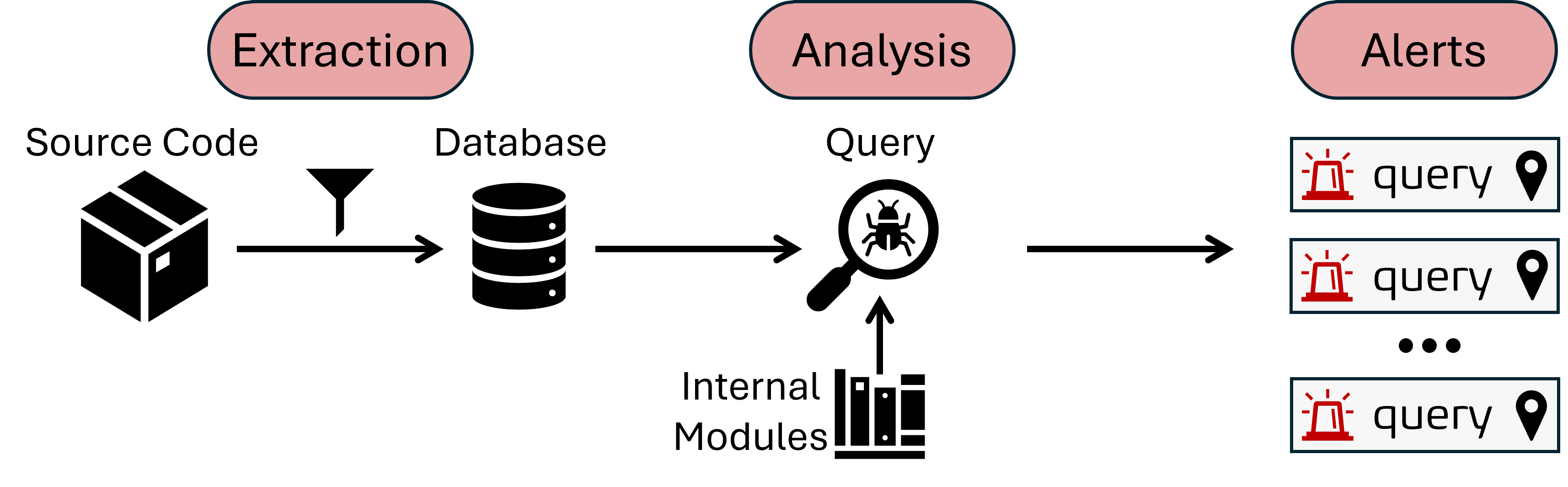}
    \caption{In a CodeQL analysis, a database is generated from the source code and queried, resulting in a set of alerts.}
    \label{fig:codeql-analysis}
\end{figure}

A CodeQL analysis (summarized in \autoref{fig:codeql-analysis}) involves three components: an extractor, a set of queries, and their underlying libraries. Given some source code, CodeQL uses an extractor of the language to convert it to a relational database (of facts about the source code) on which queries are applied to find patterns. This results in a set of alerts marked with the location (file, line, column). We detail below the components, all of which potentially changing across versions and impacting the performance of the tool. 

\shortsection{Extractors} The first step of the analysis is to map the source code to a relational database (typically done with \texttt{codeql database analyze}). To this end, language-specific extractors are provided. Their role is to extract information about the code (type information, control flow, data flow, etc.) and encode it into the database. The extractor outputs TRAP files that are used to create the database. For compiled language like C/C++, Java, or Go, the extractor is hooked into the build process to intercept calls from the compiler.

\shortsection{Queries} To run a specific analysis on the code through the extracted relational database, CodeQL provides a collection of \textit{queries}: Code in the QL language that, as their name suggests, query the database to extract relevant semantics from the code. Queries also contain valuable metadata such as their name and id, a description of what they target, expected precision of the query, the severity of the problem targeted, and tags (\eg the CWE targeted, if applicable). To simplify workflows, queries, are aggregated into query suites such as \texttt{code-scanning}, \texttt{security-extended}, \texttt{security-and-quality}. This distinction is important because it has implications on the precision of the results (which we will study in \autoref{section:evaluation-locality}). For instance, \texttt{security-extended} expands the default suite by adding lower precision and lower severity queries.

\shortsection{Libraries} Since the extracted relational database stores an enormous amount of facts about the code, creating queries from scratch every time would be inefficient and redundant. To simplify the development, libraries are developed alongside the queries to expose modules that the latter can reuse. While each language has its own collection of modules, there are a few shared libraries that rely on language-agnostic properties. For example, the module for finding cryptographic algorithms is shared between languages as it relies on recognizing a name (which is hard-coded from the names in real-world cryptography libraries).
\section{Motivation}\label{section:motivation}
\begin{figure}[h]
    \centering
    \includegraphics[width=\linewidth]{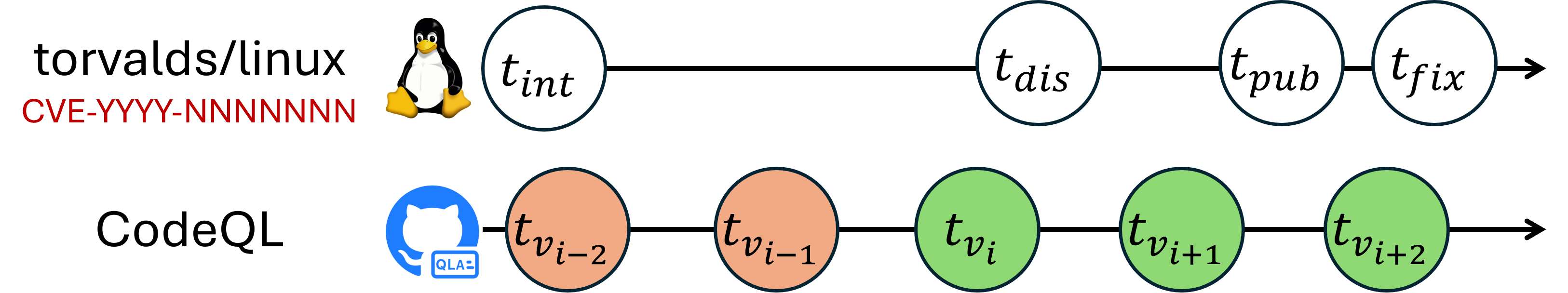}
    \caption{An example vulnerability and CodeQL timeline. A vulnerability is introduced (\(t_{\text{int}}\)), discovered (\(t_{dis}\)), then fixed (\(t_{\text{fix}}\)). In the meantime, CodeQL periodically releases new versions (\(t_{v_{k}}\)). If version \(v_{i}\) is the first to detect the vulnerability, then the efficacy of CodeQL in practice is as good as \(t_{\text{fix}} -t_{v_{i}}\).}
    \label{fig:cve-lifecycle}
\end{figure}

\shortsection{CVE lifecycle} CVEs generally follow the same lifecycle~\cite{iannoneSecretLifeSoftware2023} (as illustrated in the first row of \autoref{fig:cve-lifecycle}). They are first introduced through a vulnerability contributing commit (VCC)~\cite{hoganChallengesLabelingVulnerabilityContributing2019} (\(t_{\text{int}}\)) which can go unnoticed for a long time (\eg a 27 years vulnerability in OpenBSD~\cite{ClaudeMythosPreview}). Once discovered either by a maintainer or a security researcher (\(t_{\text{dis}}\)), they are generally fixed (\(t_\text{fix}\)) in a reasonably short amount of time. The CVE is often published (\(t_{\text{pub}}\)) at the same time as the fix (coordinated disclosure~\cite{pauleyCVEWaybackMachine2023,liuEmpiricalStudyVulnerability2025}) or before. While there are a few other steps such as the validation from the vendor (with assessment of the severity) or the coordinated disclosure, we focus on the coarse-grained timeline in this work. Along the CVE lifecycle, a SAST tool like CodeQL releases new versions periodically to improve detection capabilities and aims to detect the vulnerability as close to \(t_{\text{int}}\) as possible. Thus, measuring its effectiveness needs to take in account this temporal perspective.

\shortsection{The race to discovery} The key part of the timeline is \(t_{\text{dis}}-t_{\text{int}}\): the time it took to find the vulnerability. Indeed, the longer the duration, the more likely an adversary can find it and exploit it (\ie making it a zero-day). To this end, a lot of effort is dedicated through bug bounties~\cite{walsheEmpiricalStudyBug2020,sridharHackingGoodLeveraging2021} (\eg HackerOne~\cite{hackeroneHackerOne}), specialized security teams, and automated analyses. Among the latter, SAST tools like CodeQL aim to ease this effort by scanning the code against a knowledge base of known vulnerable code patterns. They are therefore constantly improved to help identify as many vulnerabilities as possible while limiting false positives. 

\shortsection{Problem statement} Given a CVE with the previously defined \(t_{\text{int}},\ t_{\text{dis}},\ t_\text{fix},\ t_{\text{pub}}\) and CodeQL which releases its \(k\)th version \(v_k\) at \(t_{v_{k}}\), the problem is to characterize the proactivity of CodeQL in detecting the vulnerability. In other words, assuming that there exists a CodeQL version \(v_{i}\) which detects vulnerability (\eg the version introduced a query that targets the CWE associated with the vulnerability), we want to characterize \(t_{\text{dis}}-t_{v_{i}}\).
In practice, \(t_{\text{int}}\) and \(t_{\text{dis}}\) are hard to measure at a large scale~\cite{hoganChallengesLabelingVulnerabilityContributing2019}. Further, the time window between \(t_{\text{dis}}\) and \(t_{\text{fix}}\) is highly relevant as the project associated to the CVE remains vulnerable, therein enabling a threat actor scanning for vulnerabilities (\eg using CodeQL) to exploit it. Thus, we will use \(t_\text{fix}\) as the temporal anchor and aim to characterize \(t_{\text{fix}}-t_{v_{i}}\), defined as the \textit{lead time}. Overall, by studying and quantifying this lead time, we can resolve if CodeQL automated pipelines would have helped to identify the vulnerability during its lifecycle (especially before its fix) in practice, unlike prior work that limited the study to a single point in time~\cite{bellerAnalyzingStateStatic2016,zhouComparisonStaticApplication2024}.
\section{Methodology}\label{methodology}
\begin{figure*}
    \centering
    \includegraphics[width=\textwidth]{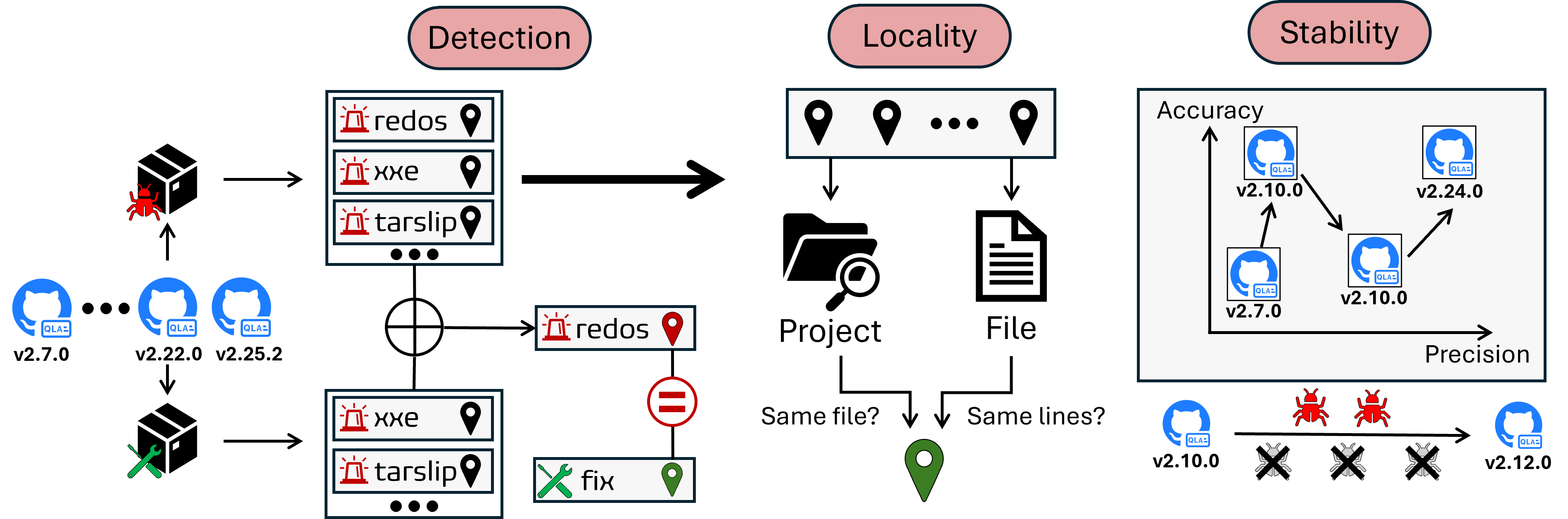}
    \caption{Overview of our methodology. We instantiate the tool (CodeQL) at a given version and apply it on two commits per CVE (the fix commit and the vulnerable commit), resulting in two sets of alerts that are compared to mark the CVE as detected or not. Then, we measure the position of alerts with respect to the vulnerability at the project and file level. Finally, we measure the movement of the tool on the accuracy-precision trade-off space and the stability of its detection across versions.}
    \label{fig:methodology-overview}
\end{figure*}

Our goal is to establish a rigorous longitudinal study over a static analyzer. Our methodology spans two key steps which we detail below: classifying a CVE as detected, and measuring the location of alerts generated by CodeQL with respect to vulnerabilities locations. \autoref{fig:methodology-overview} shows an overview of our methodology.

\shortsection{Definitions}
Let \(\mathcal{V}\) be the set of CodeQL versions and \(\mathcal{C}\) the set of CVEs. For \(cve\in\mathcal{C}\) of repository \(R_{\text{cve}}\), we define its fix commit as \(\delta_{\text{cve}} = (F_{\text{cve}},L_{\text{cve}})\) where \(F_{\text{cve}}\) is the  set of files modified and \(L_{\text{cve}}\subseteq F_{\text{cve}} \times \mathbb{N}\) is the set of (file, line) pairs of the fix.

An alert \(a\) generated by a CodeQL version \(v\in\mathcal{V}\) on a repository is a tuple \(a=(a_{\text{query}}, a_{\text{file}}, a_{\text{lines}}, a_{\text{fp}}, a_{v})\) where \(a_{\text{lines}}\) is the range of lines of code in \(f_{\text{file}}\) flagged by \(a_{\text{query}}\). In practice, the output of CodeQL in the SARIF format also reports additional locations (\texttt{relatedLocations}), but we focus on the single location reported as the alert location (\texttt{primaryLocation}). Finally, we define \(\mathcal{A}_{\text{vul}}(cve,v)\) as the set of alerts after applying CodeQL version \(v\) on the vulnerable (pre-fix) commit of \(R_{\text{cve}}\) and \(\mathcal{A}_{\text{fix}}(cve,v)\) as the set of alerts on the fix commit.

\begin{table}[ht!]
\centering
\begin{tabularx}{\columnwidth}{@{}p{2.7cm}X@{}}
\toprule
\textbf{Symbol / Term} & \textbf{Description} \\
\midrule
$\mathcal{V}$ & Set of CodeQL versions \\
$\mathcal{C}$ & Set of CVEs \\
$\delta_{\text{cve}}=(F_{\text{cve}},L_{\text{cve}})$ & Fix commit of a CVE \\
$F_{\text{cve}}$ & Files modified by the fix commit \\
$L_{\text{cve}}\subseteq F_{\text{cve}}\times\mathbb{N}$ & Pairs (file, line) mapping to fix locations \\
$a_{\text{query}}$ & Query that triggered the alert \\
$a_{\text{file}}$ & File where the alert is reported \\
$a_{\text{lines}}$ & Range of lines flagged by the alert \\
$a_{\text{fp}}$ & Fingerprint of the alert \\
$a_v$ & CodeQL version generating the alert \\
$\mathcal{A}_{\text{vul}}(cve,v)$ & Alerts on the vulnerable commit\\
$\mathcal{A}_{\text{fix}}(cve,v)$ & Alerts on the fix commit \\
$D(cve,v)$ & Predicate indicating whether $v$ detects $cve$ \\

\bottomrule
\end{tabularx}
\caption{Terminology and notation used in the methodology.}
\label{tab:terminology}
\end{table}

\subsection{Measuring Detection}\label{section:methodology-detection}
To measure the efficacy of CodeQL at a given version, we apply the tool on real-world CVEs as it shows the practical impact of the tool. In this section, we detail our selection criteria for the CVEs we apply CodeQL on, and the strategy we use to measure whether a CVE is detected given a set of alerts.

\shortsection{Sample selection} We use CVEFixes~\cite{bhandariCVEfixesAutomatedCollection2021}---a dataset of CVEs and their corresponding fix commit---as the foundation for the vulnerabilities. While the objective of this dataset was mostly to improve machine learning based approaches, its size (11\;873 CVEs corresponding to 272 CWEs) allows a more comprehensive study on real-world vulnerabilities. For a fair measurement, we keep only CVEs that were assigned an ID after the introduction of CodeQL in 2019. Then, we use the fix commit date \(t_{\text{fix}}\) as the temporal anchor: if any version of CodeQL prior to this date successfully detects the CVE, then it has a positive lead time (\(t_{v_{i}}<t_{\text{fix}}\)). While vulnerability contributing commits (VCCs)~\cite{meneelyWhenPatchGoes2013} could be a ground truth for \(t_{\text{int}}\), identifying such commits at scale remains an open challenge~\cite{hoganChallengesLabelingVulnerabilityContributing2019,baoVSZZAutomaticIdentification2022} and does not fit the goal of our analysis (as established in \autoref{section:motivation}).

\shortsection{Attributing detection} To gauge the detection capability on a CVE, we run CodeQL on repositories that have a fix commit addressing a CVE. Specifically, we run the tool on both the fix commit and the parent commit---which contains the vulnerability---and filter only the relevant CodeQL alerts following these two properties: the alerts must be (a) at the same line(s) and same file(s) as the fix, and (b) not present in the fix. The latter property is ensured using the \texttt{partialFingerprints} property from the OASIS standard~\cite{oasisStaticAnalysisResults}. Formally, a \(cve\in\mathcal{C}\) is detected by version \(v\) (noted \(D(cve,v)\)) when
\begin{equation*}
     \exists\, a\in\mathcal{A}_{\text{vul}}(c,v),\exists\, (a_{\text{file}}, l)\in L_{\text{cve}},\; 
    l \in a_{lines} \land a\not\in \mathcal{A}_{\text{fix}}(c,v).
\end{equation*}

The two heuristics work in tandem and are both necessary to exhibit CVE-related alerts as they both generate false positives for different reasons. Indeed, using location alone highly depends on the content of the fix and whether it only focuses on the vulnerability or modifies the codebase in other ways. On the other hand, selecting the set of alerts \(\mathcal{A}_{\text{vul}}(cve,v)\setminus\mathcal{A}_{\text{fix}}(cve,v)\) (present in the vulnerable commit but not in the fix commit) might yield alerts that are due to a change in the fix (not necessarily related to the CVE) that had a downstream effect on the codebase (\eg sanitizing a source could remove alerts of any matching non-sanitized sinks). 

We note that the second heuristic (filtering based on which alerts remain after the fix) assumes that the fix does remove the vulnerability. Prior work has challenged the idea that fix commits fulfill their goal and that a single fix commit is enough~\cite{bhandariCVEfixesAutomatedCollection2021,piantadosiFixingSecurityVulnerabilities2019}. However, in such event, the CVE alerts would remain after the fix. Therefore, this part of the filtering process is conservative and yields a lower bound of the true detection capabilities of CodeQL. 

To ensure that this method is accurate, we manually verified one set of alerts generated by CodeQL for each detected CVE (that verify \(D(cve,v)\)) and found that our approach to automatically label them is highly accurate with a 94\% agreement rate.
\subsection{Measuring Locality}\label{section:methodology-locality}
\begin{figure}
    \centering
    \includegraphics[width=\linewidth]{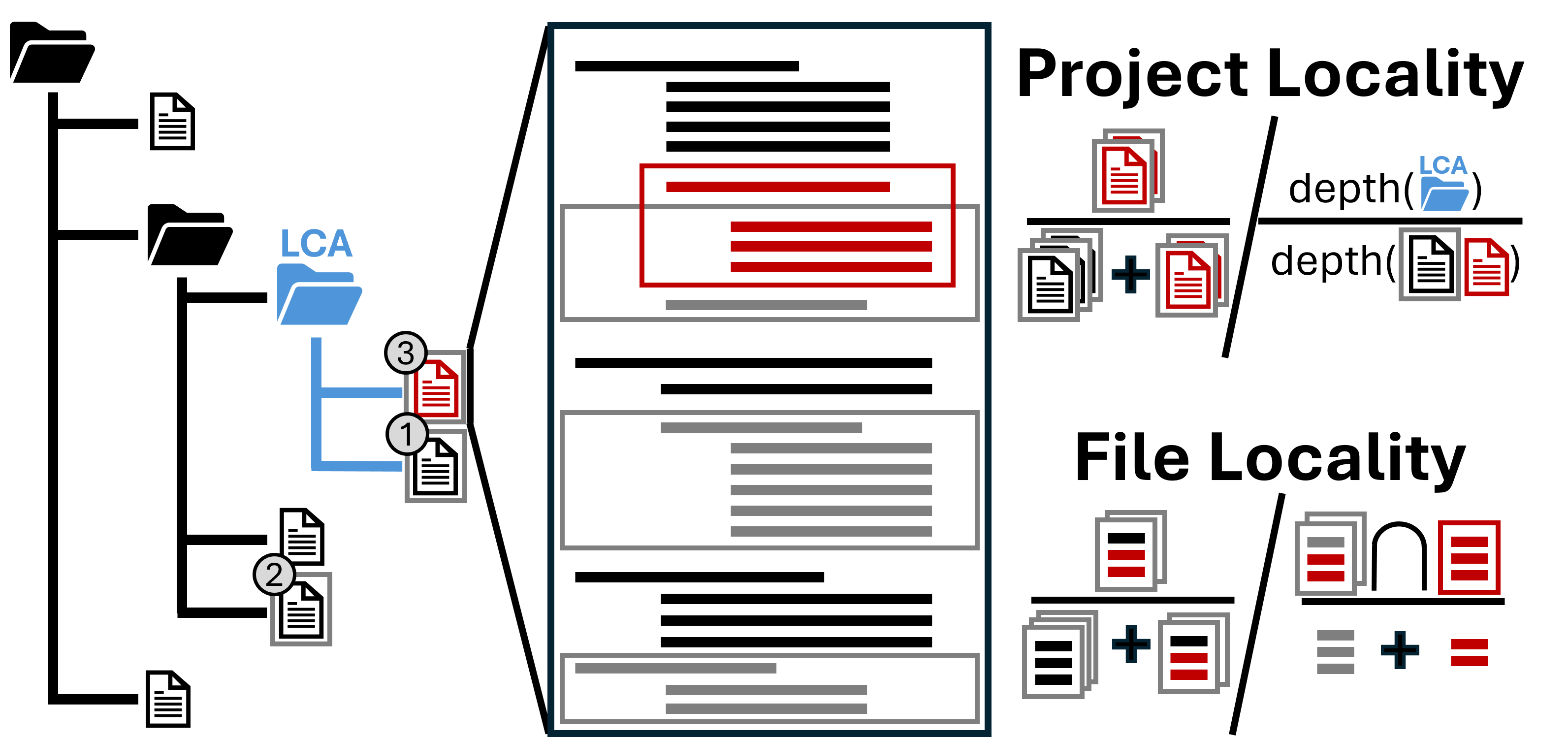}
    \caption{Overview of our measures of locality. We consider the distance between alerts and the vulnerable code at the project level (same file) and at the file level (same lines).}
    \label{fig:locality-measures}
\end{figure}

While CodeQL may detect CVEs, a detection is only meaningful if it is \textit{actionable}: the relevant alerts are localized enough for a developer or attacker to identify and act on them. Indeed, alert fatigue is a well known problem in vulnerability detection~\cite{amiFalseNegativeThat2024,johnsonWhyDontSoftware2013,christakisWhatDevelopersWant2016} and other security-related fields such as network intrusion detection~\cite{axelssonBaserateFallacyDifficulty2000}, where an abundance of alerts limits the practical use of a tool. To this end, we study the \textit{locality} of CVE detections of CodeQL, \ie the location of the alerts with respect to the vulnerable code. Using the vulnerable locations established in \autoref{section:methodology-detection} as ground truth, the locality measurement quantifies the effort required by a developer to make use of alerts generated by CodeQL. Prior work has primarily evaluated SAST tools using precision and recall metrics~\cite{ghaffarianSoftwareVulnerabilityAnalysis2017,liComparisonEvaluationStatic2023,risseTopScoreWrong2025}, without considering the spatial distribution of alerts relative to vulnerable code. Instead, we separate locality into two measurable components representing the hierarchical structure of source code: project- and file-level (visualized in \autoref{fig:locality-measures}). For each component, we consider a hard and a soft variant to represent the presence of the alert in the vulnerable locations and their distance to the vulnerable locations, respectively. We note that here, we consider all the alerts at the vulnerable commit (\(\mathcal{A}_{\text{vul}}\)) as this is what a maintainer---unaware of the vulnerability---would observe.

\shortsection{Project-level locality} In large repositories, certain regions of the code might require very narrow technical expertise (which is why concepts such as code owners dictated by the \texttt{CODEOWNERS} file matter). Therefore, when a large set of alerts is generated by a SAST, their position relative to the vulnerability is a key component for managing and triaging them. To quantify this, we use the following hard project-level locality metric which computes the proportion of generated alerts that are on a vulnerable file, \ie 
\begin{equation*}
    \frac{|\{a \in \mathcal{A}_{\text{vul}}(\text{cve}, v) \mid a_{\text{file}} \in F_{\text{cve}}\}|}{|\mathcal{A}_{\text{vul}}(\text{cve}, v)|}.
\end{equation*}

To better account for the hierarchical structure of codebases, we define a soft variant using a normalized tree similarity $\text{sim}_{\text{tree}}$. Given the vulnerable file and an alert, we first find the lowest common ancestor (LCA), as shown in blue in \autoref{fig:locality-measures}. We then compute the ratio between its depth (in the codebase) and the maximum depth between the alert's file and the vulnerable file:
\begin{equation*}
    \text{sim}_{\text{tree}}(p, q) = \frac{\text{depth}(\text{LCA}(p, q))}{\max(\text{depth}(p),\, \text{depth}(q))}.
\end{equation*}

Then, we define the soft variant of the project-level locality as the average of this similarity across generated alerts (using the closest vulnerable file if there are multiple), \ie
\begin{equation*}
    \frac{1}{|\mathcal{A}_{\text{vul}}(\text{cve}, v)|} \sum_{a \in \mathcal{A}_{\text{vul}}(\text{cve}, v)} \max_{f \in F_{\text{cve}}} \text{sim}_{\text{tree}}(a_{\text{file}}, f).
\end{equation*}

In practice, this means that we penalize alerts that are ``far away'' from the location of the vulnerability, but we still consider them. For instance, let's consider the example in \autoref{fig:locality-measures}: a total of 6 alerts scattered across 3 files, with three in the vulnerable file. Then, the hard project locality would be \(\frac{3}{6}\) while the soft counterpart is \(\frac{1}{6}(3 + \frac{2}{3} + 2\frac{1}{3}) = \frac{13}{18}\): 3 alerts in the right file (1), 1 alert in the same directory (\(\frac{2}{3}\)), 2 alerts in the parent directory (\(\frac{1}{3}\)).

\shortsection{File-level locality}
A high project-level locality does not necessarily mean that the alerts would help identify and mitigate a vulnerability. This prompts for the use of a finer-grained locality metric that focuses on the location in the code---a file locality metric---which measures the number of alerts in the vulnerable file that intersect with the vulnerability location. Formally, it is defined as
\begin{equation*}
    \frac{|\{a \in \mathcal{A}_{\text{vul}}(\text{cve}, v) \mid \exists\, (a_{\textit{file}}, l) \in L_{\text{cve}},\; l\in a_{\textit{lines}}\}|}{|\{a \in \mathcal{A}_{\text{vul}}(\text{cve}, v) \mid a_{\textit{file}} \in F_{\text{cve}}\}|}.
\end{equation*}

For a soft variant, we use the Jaccard index applied on the sets of lines. The goal here is to penalize long alerts outside the vulnerable code (that would take more effort from the developer) and instead reward overlap. The soft file-level locality is thus defined as the number overlapping lines between alerts and vulnerable code divided by the total number of unique lines across alerts and vulnerable code. Averaging over multiple files, it is formalized as
\begin{equation*}
    \frac{1}{|F_{\text{cve}}|}
    \displaystyle\sum_{\substack{ f_{\text{cve}} \in F_{\text{cve}}}} \frac{|\{ l \in a_{\text{lines}}| a\in\mathcal{A}_{\text{vul}}(cve,v) \land (l,f_{\text{cve}}) \in L_{\text{cve}} \}|}{\{l |(l,f_{\text{cve}})\in L_{\text{cve}} \lor \exists a\in \mathcal{A}_{\text{vul}}(cve,v), l\in a_{\text{lines}} \}}.
\end{equation*}

To illustrate the difference between the two file locality metrics, let us consider the example in \autoref{fig:locality-measures}: a vulnerability with four lines of code and three alerts in the vulnerable file. The hard file locality metric would yield \(\frac{1}{3}\) (one out of three alerts overlap with the vulnerability). Its soft counterpart would yield \(\frac{3}{4 + 6 + 3} = \frac{3}{13}\), therefore penalizing long alerts that missed the vulnerable code.

\shortsection{Query suites} CodeQL ships with multiple ``query suites'': predefined sets of queries. While the default suite is generally sufficient, the \texttt{security-extended} suite allows the detection of more potential issues at the cost of precision. Therefore, we will take in consideration those two suites for the locality as they represent the choice of a developer between fewer alerts (default suite) and more thorough analyses (extended suite). 
The \texttt{security-and-quality} suite, a superset of the \texttt{security-extended} suite, is out-of-scope as it adds queries on the maintainability and reliability of the code.
\begin{table*}[h]
\centering
\small
\caption{Overview of the measurement. We select CVEs from CVEfixes~\cite{bhandariCVEfixesAutomatedCollection2021} between 2019 and 2024 across six languages. We then remove CVEs where all CodeQL versions output an error, and identify the CVEs detected by CodeQL (following the criterion from \autoref{section:methodology-detection}). We additionally report the number of unique CWEs, total lines of code, and total number of alerts generated.}
\begin{tabular}{l c c c c c c c c c c}
\toprule
 & Total & \multicolumn{6}{c}{Languages} & & & \\
\cmidrule(lr){3-8}
 & \#CVEs (\#repos) & C/C++ & Go & Java & JavaScript & Python & Ruby & \#CWEs & \#LoC & \#Alerts \\
\midrule
\textbf{Total} & \textbf{3,993 (1,622)} & \textbf{2,094 (495)} & \textbf{359 (232)} & \textbf{315 (172)} & \textbf{694 (464)} & \textbf{353 (229)} & \textbf{178 (82)} & \textbf{177} & \textbf{20,117,771,484} & \textbf{2,768,244} \\
2019 & 321 (143) & 261 (95) & 8 (8) & 14 (10) & 27 (23) & 7 (7) & 4 (4) & 69 & 4,911,729,711 & 140,708 \\
2020 & 403 (240) & 240 (104) & 27 (22) & 28 (23) & 75 (65) & 18 (17) & 15 (13) & 87 & 1,662,775,560 & 149,481 \\
2021 & 730 (364) & 446 (132) & 41 (34) & 45 (30) & 131 (116) & 50 (43) & 17 (13) & 98 & 2,310,103,776 & 525,704 \\
2022 & 1,174 (521) & 658 (152) & 101 (81) & 104 (70) & 181 (145) & 79 (60) & 51 (26) & 124 & 5,809,386,700 & 929,251 \\
2023 & 918 (469) & 350 (127) & 111 (81) & 91 (48) & 196 (123) & 115 (77) & 55 (24) & 117 & 4,222,296,925 & 737,723 \\
2024 & 447 (292) & 139 (80) & 71 (51) & 33 (24) & 84 (59) & 84 (64) & 36 (21) & 55 & 1,201,478,812 & 285,377 \\
\midrule
\makecell[l]{Successful\\analyses} & 2,440 (1,166) & 893 (203) & 289 (180) & 68 (59) & 685 (456) & 337 (219) & 168 (75) & 160 & 1,603,436,221 & 2,766,274 \\
\midrule
\textbf{Detected} & \textbf{171 (150)} & \textbf{15 (10)} & \textbf{18 (14)} & \textbf{11 (11)} & \textbf{79 (69)} & \textbf{36 (34)} & \textbf{12 (12)} & \textbf{40} & \textbf{31,426,616} & \textbf{859,914} \\
\bottomrule
\end{tabular}
\label{tab:cohort-overview}
\end{table*}

\section{Evaluation}\label{eval}
We apply our approach to answer the following research questions:

\begin{rqlist}[series=rquestion]
    \item What CVEs does CodeQL identify and starting when?\label{rq1}
    \item Among detected CVEs, were alerts located near the vulnerability location?\label{rq2}
    \item Is the effectiveness of CodeQL stable across versions?\label{rq3}
\end{rqlist}

\subsection{Experimental Setup}
All experiments where CodeQL was applied on a set of CVEs were run on a cluster. Each run was set up in a Docker container with 64~GB of memory, 50~GB of disk space, and 2 CPU cores. The total runtime for all experiments amounts to more than 17\;376 hours (724 days) for a total of more than 30B lines of code analyzed across versions and 2.7M alerts generated (see overview in \autoref{tab:cohort-overview}).

\shortsection{Time frame and versions} The temporal frame in this study starts at the introduction of CodeQL as an open-source tool (v2.0.0 released on November 14, 2019) and ends at the latest available version at the time of the study (v2.25.2, released in April 2026). Therefore, we filter for CVEs that were assigned in 2019 or later. Since CodeQL spans many versions (129 publicly observable from v2.4.0 to v2.25.2), we break down the analysis in two phases: we first apply the entire corpus (\NumCVEs CVEs) every 2 minor version (v2.8.0, v2.10.0, etc.), then we run the remaining 105 versions (between v2.7.0 and v2.25.2) on all CVEs that were detected by at least one of the 9 versions under the assumption that a negligible amount of CVEs would only be detected by the versions in-between. We note that versions prior to v2.4.0 are not properly documented (\eg no changelogs), likely due to the transition from Semmle to GitHub. Further, we observed many breaking changes in versions prior to v2.7.0 leading to a significant number of failed analyses. Therefore, we left these versions out of scope of this study to ensure that they are evaluated on the same corpus.

\shortsection{CLI and queries} CodeQL comes with two user-facing components: a command-line interface (CLI) spanning more than 150 versions, available as binaries on GitHub (\url{https://github.com/github/codeql-cli-binaries}), and the set of queries and underlying libraries that power the CLI, open-sourced in a GitHub repository~\cite{githubGithubCodeql2026}. To instantiate a version of the tool, we download the release of the CLI at the version, clone the GitHub repository of the CodeQL queries and libraries, and checkout to the commit tagged with the version. We then run the \texttt{database create} and \texttt{database analyze} commands using the \texttt{security-extended} query suite of each language (which includes the default suite).

\shortsection{Repositories and environments} We focus on six languages: C/C++, Java, JavaScript, Go, Python, and Ruby. This leads to a total of 1\;622 unique repositories, each repository having its own dependencies and environment (\eg version of the language, compiler, etc.). To ensure that the measurement is functional in most cases, we set up environments that enable CodeQL analyses on all repositories. For compiled languages like Java, CodeQL needs to build the code to extract the program semantics. Therefore, we create one Docker image per language and  across multiple versions to maximize compatibility.  During the extraction phase, CodeQL attempts to automatically build the project through the \texttt{autobuild} script by default. However, this default script does not scale to our corpus of 1\;622 repositories which may span different versions of C/C++, Go, or Java. Thus, we create custom scripts that maximize compatibility across all settings. We note that CodeQL introduced the \texttt{build=none} parameter that allows the analysis without a build. As it was introduced fairly late (after v2.23.3 for C/C++), we consider it out-of-scope but acknowledge that this improves the usability of CodeQL. In practice, a maintainer of a project would provide the right environment for the analysis and the correct build command.
 

\subsection{Detected CVEs}\label{section:evaluation-detection}
In this section, we focus on measuring the effectiveness of CodeQL and answer \ref{rq1}: \textit{is CodeQL effective at detecting vulnerabilities and, if so, starting at which version?} To answer this, we first identify which CVEs can be detected by CodeQL. Then, we examine which languages and CWE categories are most represented among detections, paired with their severity level, to find the conditions and impact of detection. Next, we analyze the lead time (the duration \(t_{\text{fix}} - t_{v_{i}}\) between the fix commit of a CVE and the first version that detects the vulnerability) and proactivity of queries from detections to determine how early CodeQL identifies vulnerabilities. To this end, we synthesize our findings into a best practice for preventing vulnerabilities in OSS.

\shortsection{From all CVEs to detected CVEs} To first answer the question, we need to identify the set of CVEs detected at some point by CodeQL. \autoref{tab:cohort-overview} shows an overview of the analysis, including the number of CVEs, CWEs, and repositories. We see that except for C/C++ and Java where 50\% to 80\% of the CVEs could not be analyzed due to build constraints, our environment setup was successful in running the analysis for the other languages. Out of the \NumCVEsRan CVEs that were analyzed, \NumDetectedCVEs were detected. While this number could be seen as low (about 10\%), it is significant as OSS projects are heavily scrutinized and their vulnerabilities are complex while CodeQL remains a best effort at characterizing entire classes of vulnerabilities across many languages while limiting false detections. Further, a vulnerability of an OSS project can have a drastic impact on the security of entire ecosystems~\cite{eversonLog4shellRedefiningWeb2022}, making any correct detection crucial.

\shortsection{Languages and detection} Among detected CVEs, we see a striking 79 CVEs detected for JavaScript and 36 for Python, far above the next language with the most detected vulnerabilities. Conversely, compiled languages like C/C++, or Java exhibit lower successful analyses rate and therefore fewer detected CVEs. As newer CodeQL versions add the \texttt{build=none} mode (which removes the need for building the project) for more and more compiled languages, this difference between languages is more likely to diminish with the caveat that the extraction might be less precise. Another factor is the inherent bias of CWE type across languages: CodeQL's queries are overwhelmingly built around source-to-sink dataflow tracking, thus the detection rate is driven by the proportion of vulnerabilities corresponding to specific CWEs.

\begin{figure}[h]
    \centering
    \includegraphics[width=\linewidth]{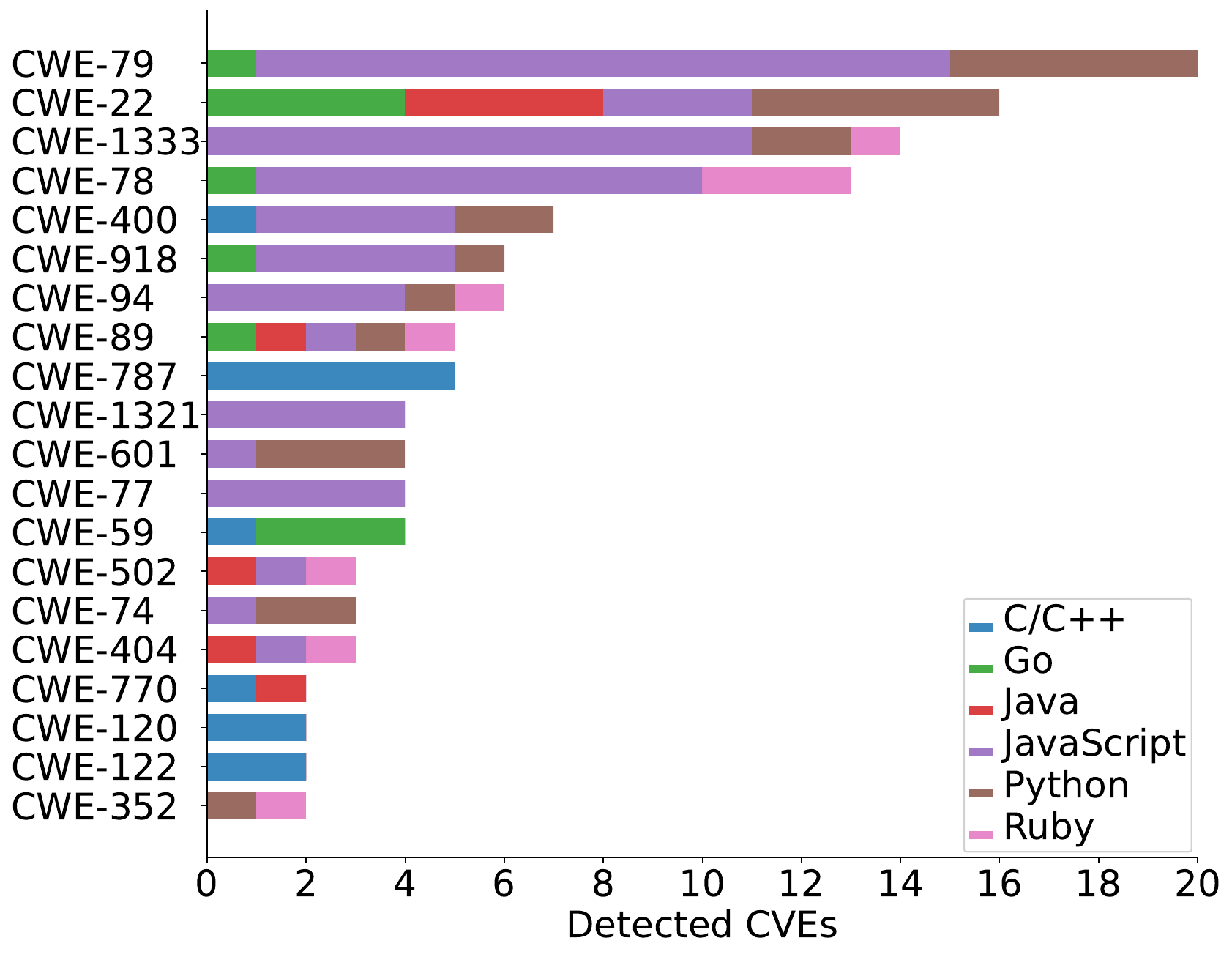}
    \caption{Top 20 CWEs detected across languages}
    \label{fig:cwe-languages}
\end{figure}

\shortsection{CWEs} Security queries are built to scan for CWEs and not specific vulnerabilities (with some exceptions\footnote{\url{https://github.com/github/codeql/tree/main/python/ql/src/Security/CVE-2018-1281}}). However, some CWEs are harder to detect than others: an SQL injection (CWE-089) is easier to detect than memory management weaknesses like Use After Free (CWE-416) as the latter might only be triggered in very specific conditions. \autoref{fig:cwe-languages} shows the top 20 CWEs detected across languages. First, the two best performing CWEs, CWE-79 (Cross-site Scripting) and CWE-22 (Path Traversal) account for nearly 36 of the detected CVEs. This result is very consistent with prior work~\cite{liIRISLLMAssistedStatic2025} and as expected for CWE-22 which is a classical taint analysis problem that static analysis tools like CodeQL are effective at. All those results are all the more important that 7 of the CWEs in \autoref{fig:cwe-languages} are in the 2025 CWE Top 25 Most Dangerous Software Weaknesses from MITRE~\cite{CWE2025CWE}, showing an alignment between the tool's development and what is prioritized in the space.

\begin{table}[h]
  \centering
  \begin{tabular}{lrrrrr}
    \toprule
    Cohort & \#CVEs & Low & Medium & High & Critical \\
    \midrule
    All CVEs & 3993 & 281 & 1631 & 1512 & 565 \\
    Detected & 171 & 14 & 68 & 63 & 26 \\
    Positive lead time & 83 & 4 & 42 & 25 & 12 \\
    \bottomrule
  \end{tabular}
  \caption{CVSS severity distribution across CVEs.}
  \label{table:cohort-severity}
\end{table}

\shortsection{Severity} SAST tools like CodeQL are explicitly geared toward mitigating the most severe vulnerabilities to avoid overwhelming developers with alerts on low severity vulnerabilities, even if they are correct. \autoref{table:cohort-severity} shows the count of CVEs across severities according to their CVSS~\cite{first.orginc.CVSSV31Specification} score and classification. From the table, it is clear that low severity vulnerabilities are not a priority compared to high and critical severities, which is consistent with their goal. Importantly, among the CVEs that would have been detected in practice, 12 of them were assigned a critical severity, which emphasizes the importance of using these tools and leads to the following best practice for open-source software.

\begin{figure}[h!]
    \centering
    \includegraphics[width=\linewidth]{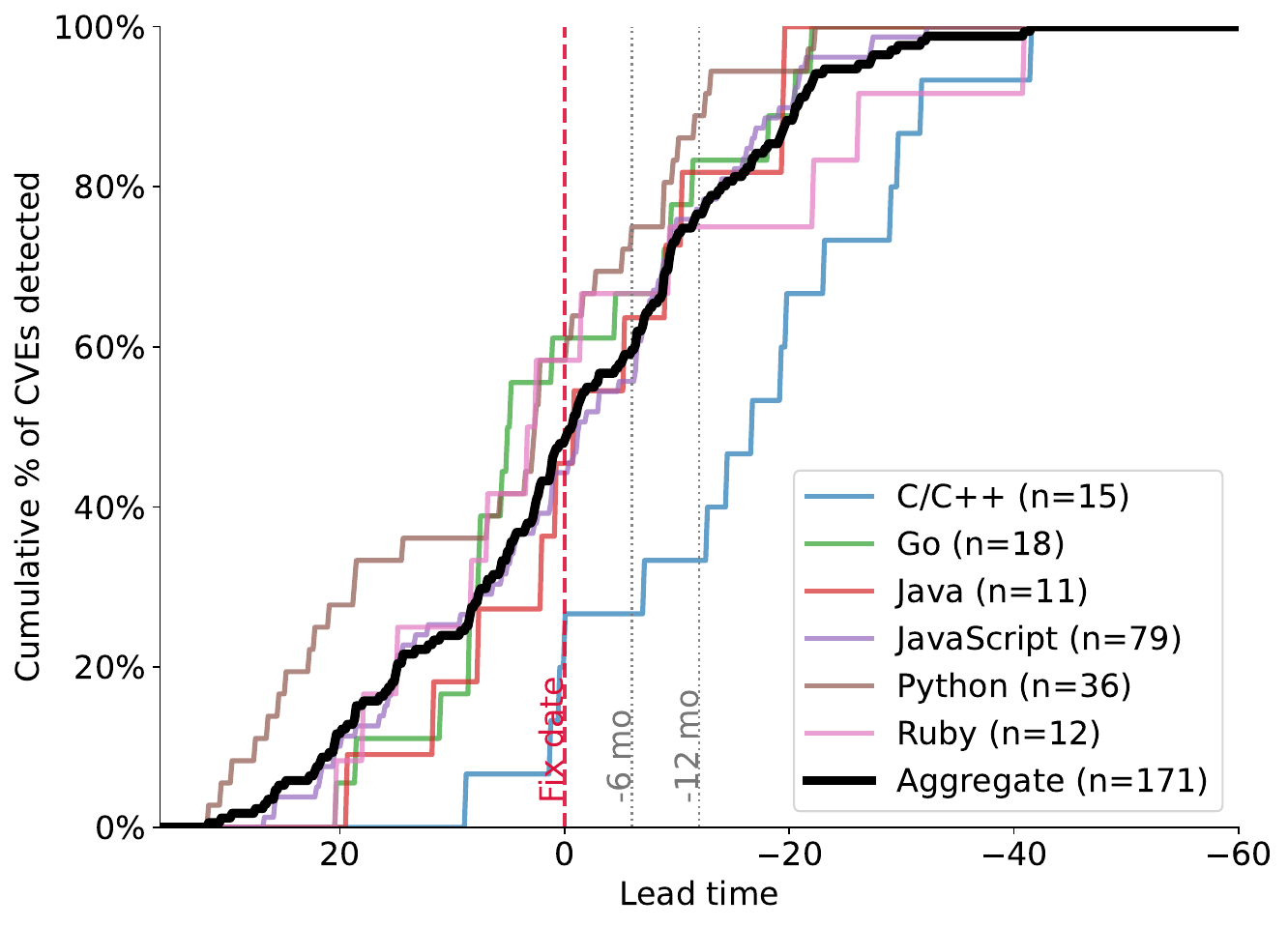}
    \caption{Lead time of detecting CVEs}
    \label{fig:lead-time-cdf}
\end{figure}

\shortsection{Detection and lead time} As stated in \autoref{section:motivation}, the main problem is not the detection of vulnerabilities, but starting which version CodeQL can detect them. Therefore, we study the lead time \(t_{\text{fix}} - f_{v_{i}}\) (as defined in \autoref{section:motivation}). Indeed, given our longitudinal measurement, we can establish a precise characterization of this lead time that prior evaluations missed, therein painting a more accurate picture of the effectiveness of the tool. \autoref{fig:lead-time-cdf} shows the CDF of CVEs detection with respect to the lead time. Overall, we note that among the \NumDetectedCVEs CVEs detected, \PercentDetectedCVEsInTime (\NumDetectedCVEsInTime) would have been identified with a version released available before the fix date. This result encourages repositories to make use of CodeQL as part of the pipeline to mitigate the introduction of vulnerabilities. While the trend is similar across languages, we note that C/C++ define the lower bound of the CDFs while Python defines the upper bound. This is surprising because C and C++ were among the first languages supported by CodeQL alongside Java. We attribute this to the amount of repositories that had an analysis failure as C/C++ exhibited one of the highest rate of build failures (57\%), making it the most challenging language to study.

\begin{tcolorbox}
    \shortsection{\result{}: Lead time} Among \NumDetectedCVEs detected CVEs, \NumDetectedCVEsInTime could have been detected by CodeQL before their fix.
\end{tcolorbox}

\begin{figure}[h]
    \centering
    \includegraphics[width=\linewidth]{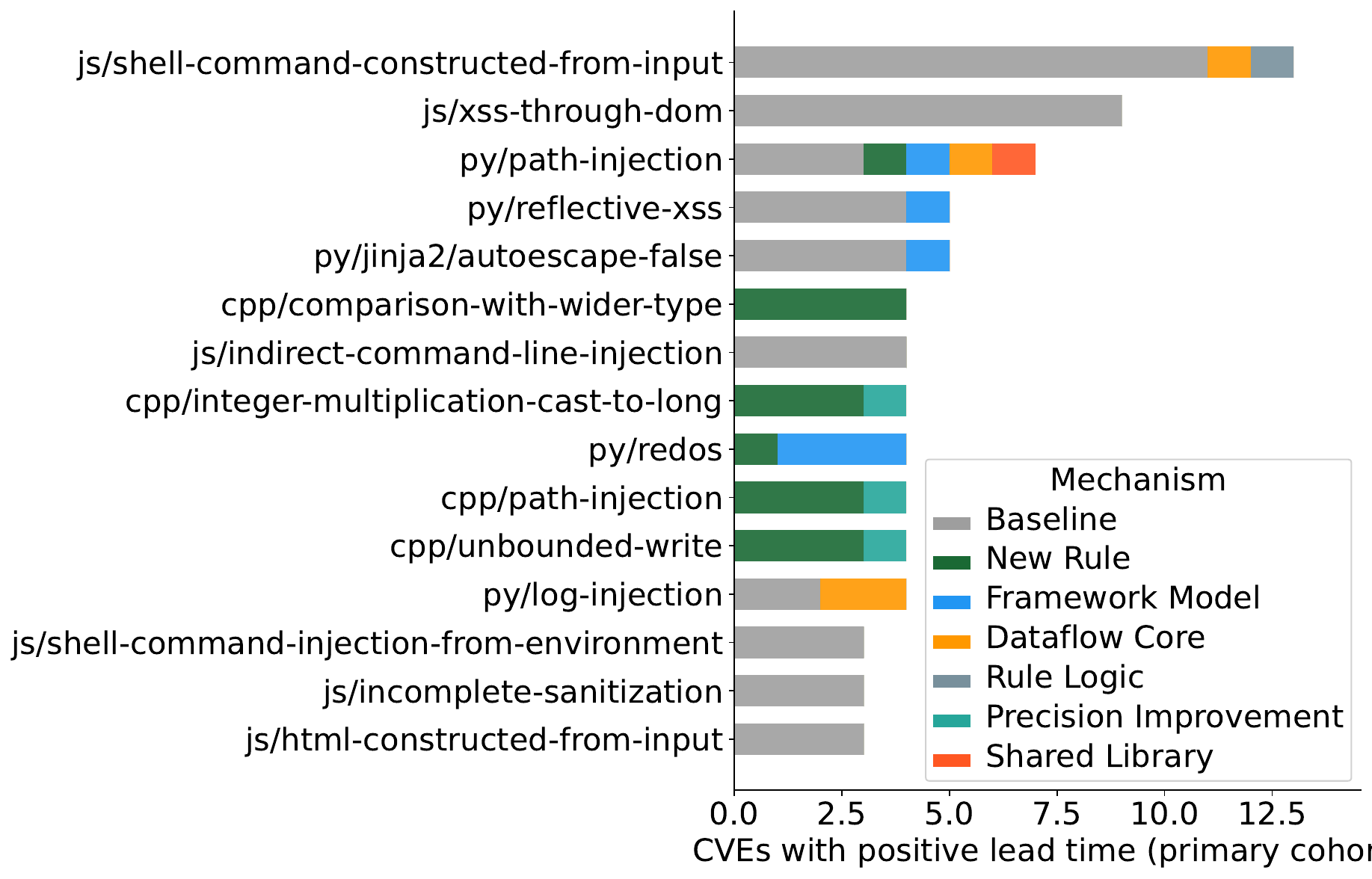}
    \caption{Most proactive queries and their most likely driving mechanism. Baseline represents earliest detections (v2.7.0) from which we cannot conclude in any mechanism.}
    \label{fig:proactive-changes}
\end{figure}

\shortsection{Proactivity of queries} Queries are the main drivers of the analyses and are thus one the most important components to consider. As CodeQL (formerly LGTM) had a substantial amount of development before being public, it started out with a large amount of queries, hence the large amount of detections in v2.7.0. \autoref{fig:proactive-changes} reports the top 15 queries by number of CVEs with a positive lead time, along with the most important change between the version that detected the CVE and the previous version (obtained through the changelogs). We remark that the results echo the earlier CWE results from \autoref{fig:cwe-languages} since \texttt{xss-through-dom} maps to CWE-79 (Cross-Site Scripting) and \texttt{path-injection} maps to CWE-22 (Path Traversal). Beyond the high amount of baseline detections (\ie at the first version analyzed, v2.7.0), we see that while the predominant mechanism is the introduction of the query itself, some low-level changes (\eg the core dataflow library) increase the coverage of CVEs correctly identified.

\shortsection{Adoption in vulnerable repositories and adversarial utility} We performed a follow-up study on the adoption of CodeQL by the repositories linked to the CVEs we analyzed. As reported in \autoref{section:background}, CodeQL is a popular tool (up to 30\% adoption as seen in \autoref{fig:codeql-adoption}). We found that 1,196 (71\%) of the 1,622 unique repositories mapping to CVEs never enabled CodeQL, and 185 only had it after the vulnerability fix. More importantly, for 69 of the CVEs that could have been detected prior to the fix, the corresponding repositories never enabled CodeQL. As explored by prior work, the main factor that prevents developers from using SAST tools is the number of false positives~\cite{johnsonWhyDontSoftware2013}, which incurs a cost for them and lead to alert fatigue (which we will tackle in more details in \autoref{section:evaluation-locality}). This is concerning as SAST tools can also be used by adversaries to identify vulnerable code. Indeed, while cutting-edge advances in cybersecurity capabilities of large language models warrant for restrictions on usage~\cite{ClaudeMythosPreview}, CodeQL is available to anyone, including adversaries. Therefore, repositories that do not enable CodeQL are exposed to threat actors using it or a more advanced tool. Thus, it is crucial for OSS to integrate such tools to their pipeline, leading to the first best practice below.

\begin{tcolorbox}[colback=red!5!white, colframe=red!75!black]
  \textbf{Best Practice 1: Prevention.} SAST tools should be part of OSS pipelines to prevent the introduction of vulnerabilities.
\end{tcolorbox}

\shortsection{Wall of fame} We cross-referenced the CVEs with the CodeQL Wall of Fame~\cite{githubCodeQLWallFame}: a list of vulnerabilities that have been discovered using CodeQL by security researchers. From there, we found that 44 of such vulnerabilities overlap with our corpus, with 20 of them considered detected with our measurement. The 24 remaining can be attributed to a second use case of CodeQL beyond automated checks: helping security researchers (often from GitHub Security Lab) search for complex semantic-level patterns in the code. This illustrates that while predefined queries are helpful in detecting some vulnerabilities, they remain a best-effort at generalizing detection while remaining precise. Thus, the two main use cases of CodeQL---automating checks and helping security researchers---sit at opposite regions of the accuracy-precision trade-off: the former focuses on being as precise as possible at the cost of fewer detections while the latter maximizes the number of detections through less precise, dedicated queries. While we have seen in this section that CodeQL is effective along both use cases and that it is best practice to use it, its effectiveness is only as good as the actionability of the alerts it generates, leading us to our next research question.
\subsection{Actionability of Detections}\label{section:evaluation-locality}
After establishing the effectiveness of CodeQL in detecting vulnerabilities before their fix, we ask \ref{rq2}: \textit{are the alerts related to vulnerabilities detected by CodeQL actionable?} That is, are the detections close enough to the vulnerability for a developer to act on them. We begin our analysis by collecting the pre-fix alerts of the detected vulnerabilities to assess their locality at the project and file levels using our metrics defined in \autoref{section:methodology-locality}. With locality of alerts established, we characterize the tool in practice by analyzing how different query suites affect locality and efficiency, concluding on community best practices for triaging CodeQL alerts.

\begin{table}[htbp]
\centering
\small
\caption{Distribution of pre-fix alert counts (\(|\mathcal{F}_{\text{vul}}(cve,v)|\)).}
\label{tab:alerts_distribution}
\begin{tabular}{lrcccc}
\toprule
Group & $n$ & P25 & P50 & P75 & P90 \\
\midrule
All CVEs (clean build) & 2441 & 2 & 14 & 72 & 280 \\
Detected CVEs & 171 & 5 & 14 & 52 & 124 \\
Positive lead time & 83 & 7 & 22 & 56 & 123 \\
\bottomrule
\end{tabular}
\end{table}

\shortsection{Number of alerts} To further motivate the need for measuring actionability, we report in \autoref{tab:alerts_distribution} the distribution of the number of alerts on the vulnerable commit across settings. While for 50\% of cases it is rare to exceed 20 alerts, larger projects can generate hundreds of alerts. From a developer's perspective, this number represents one of the many tasks to do (adding features, going through issues and pull requests, fixing bugs, \etc). If those alerts are not close to the vulnerable location, the developer is less likely to find and fix the vulnerability. Using the CVEs detected by CodeQL, we can perform a retrospective on how actionable the alerts for these CVEs were through the locality measures introduced in \autoref{section:methodology-locality}.

\begin{figure}[h!]
    \centering
    \begin{subfigure}[b]{.49\columnwidth}
        \centering
        \includegraphics[width=\textwidth]{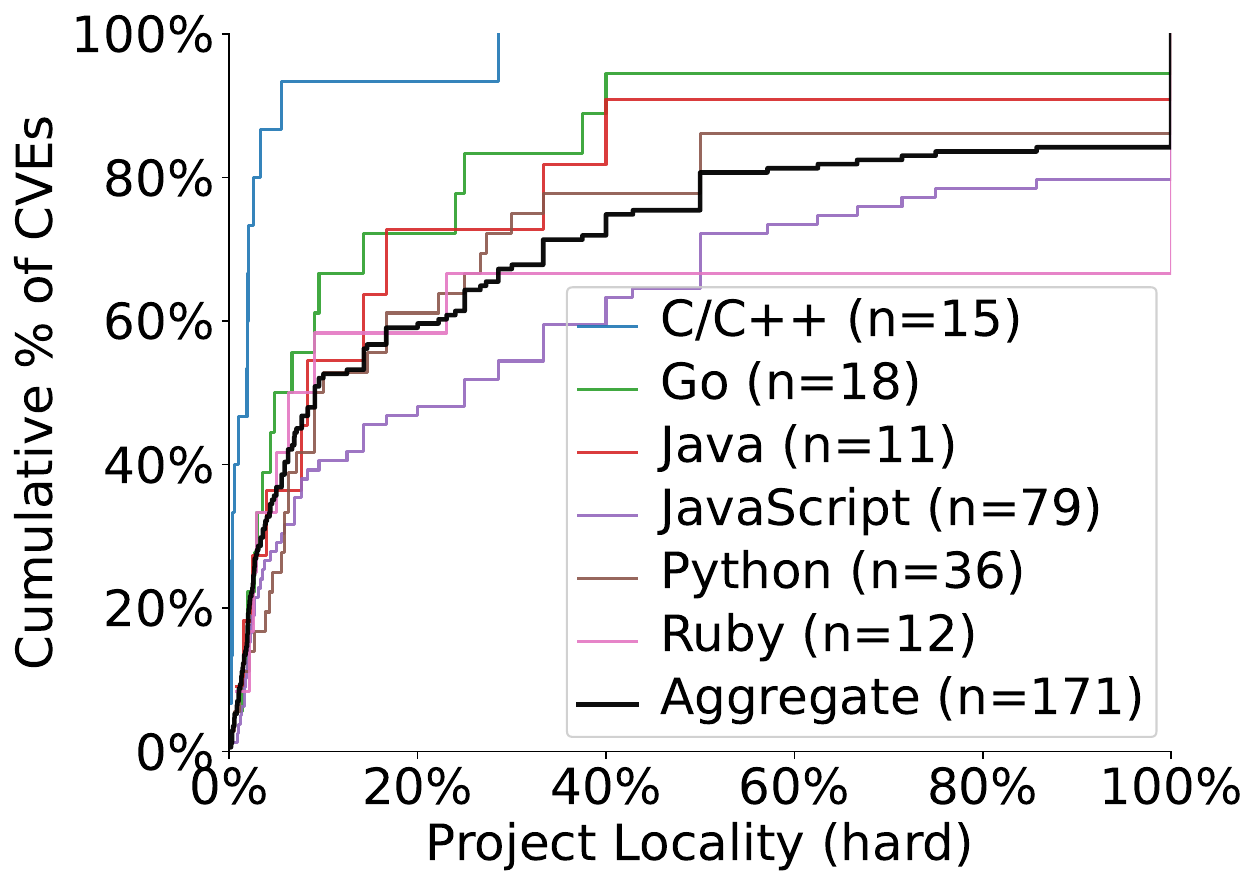}
        \caption{\textit{Hard}}
        \label{subfig:project-locality-hard}
    \end{subfigure}
    \begin{subfigure}[b]{.49\columnwidth}
        \centering
        \includegraphics[width=\textwidth]{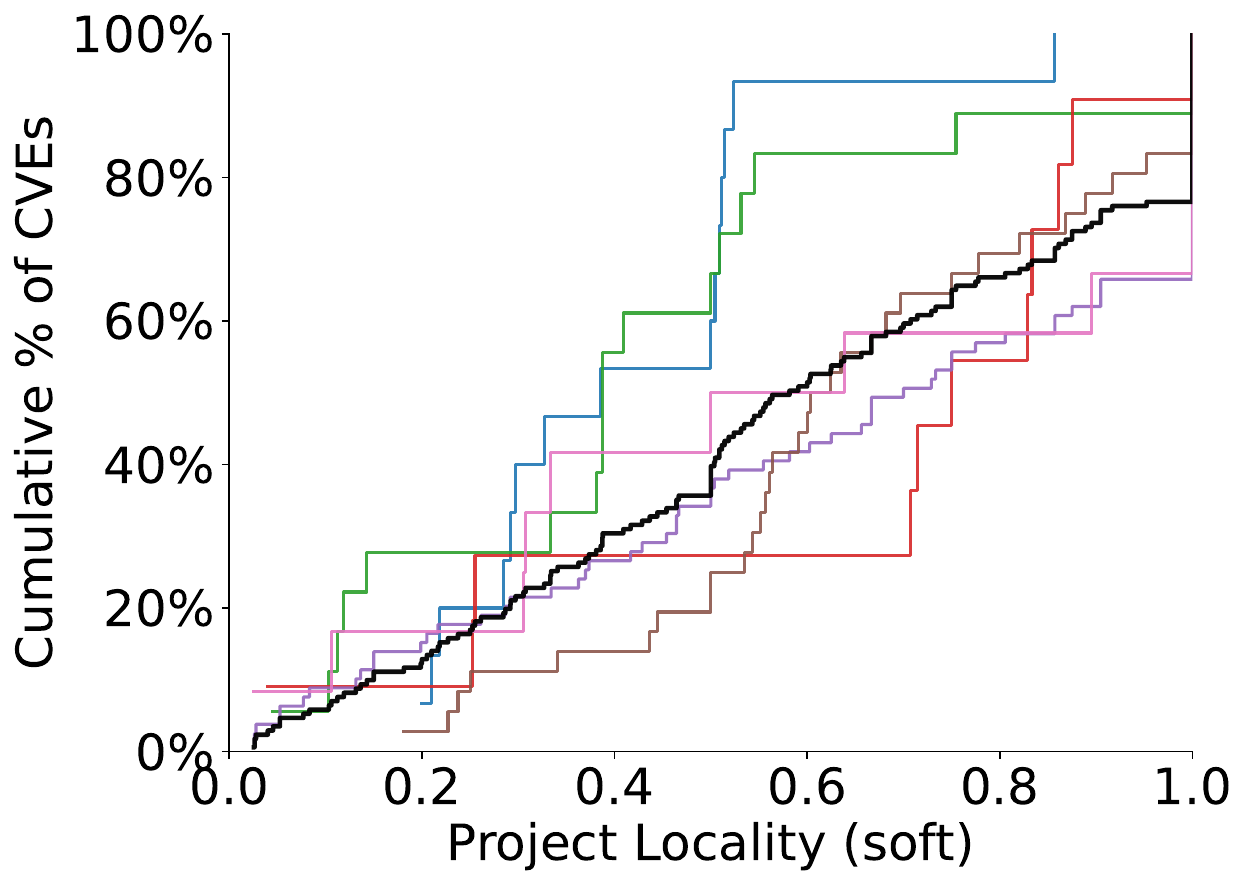}
        \caption{\textit{Soft}}
        \label{subfig:project-locality-soft}
    \end{subfigure}
    \caption{Project locality}
    \label{fig:project-locality}
\end{figure}

\shortsection{Project locality} CodeQL is generally applied on the whole repository, thus a set of alerts can span many files unrelated to the CVE and therefore make it harder to triage and fix. \autoref{fig:project-locality} shows a CDF of the hard and soft project locality metrics across languages. From the hard metric, we see that the median percentage of alerts in the right file is close to 10\%, which is fairly low. Using the soft locality metric that considers hierarchy, we obtain a fairly linear pattern of the CDF for more than 70\% of CVEs, indicating a uniform distribution of this metric. In practice, this means that alerts that contain a vulnerability detection are as likely to be close to the vulnerability than to be far from it. Those two results indicate that alerts are generally scattered across the codebase, making the triage at this level the most important step.

\begin{figure}[h!]
    \centering
    \begin{subfigure}[b]{.49\columnwidth}
        \centering
        \includegraphics[width=\textwidth]{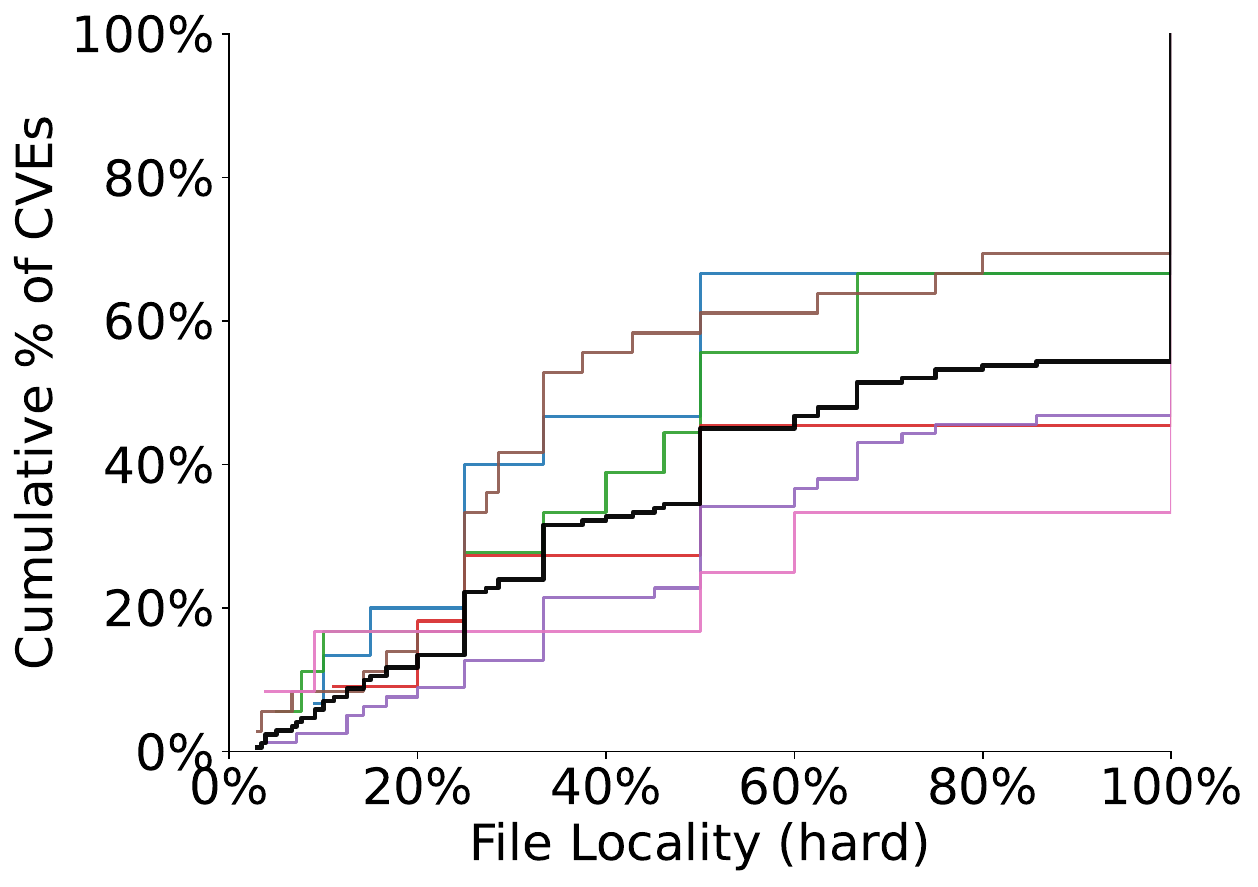}
        \caption{\textit{Hard}}
        \label{subfig:file-locality-hard}
    \end{subfigure}
    \begin{subfigure}[b]{.49\columnwidth}
        \centering
        \includegraphics[width=\textwidth]{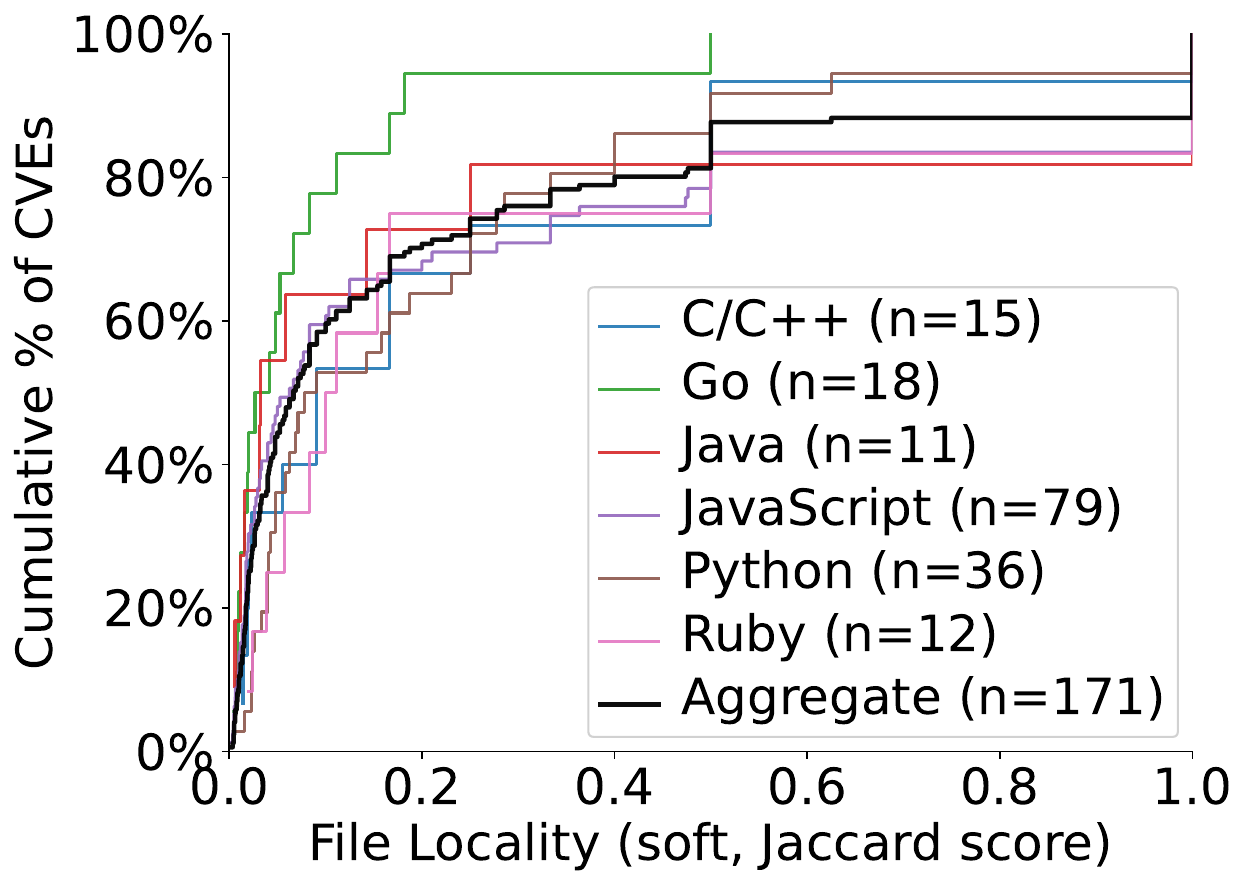}
        \caption{\textit{Soft}}
        \label{subfig:fille-locality-soft}
    \end{subfigure}
    \caption{File locality}
    \label{fig:file-locality}
\end{figure}

\shortsection{File locality} Even if the alerts span the vulnerable file, it is unclear whether they overlap with vulnerable lines of code. \autoref{fig:file-locality} shows that given the vulnerable file, alerts are very precise: 50\% of alerts are within the vulnerable location in the majority of cases. Using the soft file locality which measures more accurately the average overlap of alerts lines and vulnerable lines, we see that the overlap is generally weak with a median of 0.05. In practice, this means that provided the vulnerable region of the codebase to observe (\eg the code that handles user inputs), alerts are in generally close to the vulnerability, but do not overlap with it.

\begin{tcolorbox}
    \shortsection{\result{}: Locality} Alerts of detected CVEs are generally scattered across the codebase, but very close to the vulnerable lines given the vulnerable file.
\end{tcolorbox}

\shortsection{Query suites and reported precision} Most queries come with a precision value (either \texttt{low}, \texttt{medium}, \texttt{high}, and rarely \texttt{very-high}), which also determines their query suite. \autoref{tab:tier_locality} shows for each reported precision the number of detected CVEs, and the median locality of their generated alerts. We note that the reported precision matches well what is expected: queries reported as highly precise achieve twice the file locality. This validation reinforces that such metadata can be trusted to triage alerts and focus on the most likely weaknesses in the code.
As explained in \autoref{section:methodology-locality}, we consider both the default \texttt{code-scanning} and the \texttt{security-extended} suites which can be selected when initializing CodeQL. Therefore, we can verify the importance of this selection when it comes to locality and usefulness. Overall, we found that extending the suite did result in a much lower locality values, but given the increase the number of detected CVEs by about 70, the price to pay in locality (thus precision) is fairly low.

\begin{table}[h]
\centering
\caption{Median and standard deviation of locality metrics by reported precision of the queries.}
\label{tab:tier_locality}
\begin{tabular}{lccccc}
\toprule
\multirow{2}{*}{Tier (CVEs)} & \multicolumn{2}{c}{File locality} & \multicolumn{2}{c}{Project locality} \\
\cmidrule(lr){2-3} \cmidrule(lr){4-5}
 & Hard (\%) & Soft & Hard (\%) & Soft \\
\midrule
High (100) & $100 \pm 35$ & $6.6 \pm 17.3$ & $14 \pm 37$ & $62.5 \pm 32.3$ \\
Both (32) & $48 \pm 30$ & $6.1 \pm 14.5$ & $13 \pm 23$ & $67.5 \pm 29.2$ \\
Medium (37) & $33 \pm 35$ & $5.7 \pm 14.3$ & $5 \pm 37$ & $60.3 \pm 26.8$ \\
\bottomrule
\end{tabular}
\end{table}

\begin{table}[h!]
  \centering
    \caption{CodeQL runtime (seconds) percentiles per language for detected and all successful runs (in parentheses).}
  \label{tab:codeql-runtime}
  \begin{tabular}{lrrrr}
    \toprule
    Language & P25 & P50 & P75 & P90 \\
    \midrule
    C/C++ & 102 (94) & 425 (183) & 805 (708) & 1439 (1695) \\
    Go & 171 (100) & 316 (279) & 605 (772) & 2303 (2909) \\
    Java & 186 (182) & 411 (411) & 882 (989) & 1554 (2013) \\
    JavaScript & 64 (67) & 86 (96) & 135 (167) & 259 (328) \\
    Python & 79 (69) & 147 (132) & 329 (261) & 702 (591) \\
    Ruby & 37 (38) & 55 (68) & 99 (173) & 243 (332) \\
    \bottomrule
  \end{tabular}
\end{table}

\shortsection{Efficiency and adversarial effort} Even if a tool is effective at detecting vulnerabilities, it comes with a cost in time and resources. \autoref{tab:codeql-runtime} shows the runtime across languages at different percentiles. First, we see that for interpreted languages (JavaScript, Python, Ruby), the runtime is in more than 90\% of cases under 10 minutes which is acceptable~\cite{hiltonTradeoffsContinuousIntegration2017}. On the other hand, compiled languages exhibit a high range of runtime values: the majority of runs take less than 10 minutes but the runtime can approach an hour for the 10\% longest runs. While this result would hardly scale for that need to run a CodeQL workflow often (\eg for every pull request review), it remains reasonable given the protection that it can give. Moreover, there has been work on reducing the runtime of the analyses through incrementalization~\cite{szaboIncrementalizingProductionCodeQL2023} (only analyzing new or changed code) which started being incorporated in pull requests automated analyses starting May 2025~\cite{githubIncrementalSecurityAnalysis2025} and improved recently~\cite{githubFasterIncrementalAnalysis2026}. These improvements reinforce the importance of the previous best practice as open-source projects should be able to use the tool more often. Additionally, as CodeQL can be leveraged by malicious actors to identify flaws to exploit on OSS, efficiency gains of the tool become an opportunity for these actors to scale up the identification. Thus, for repositories using CodeQL, the following best practice to triage alerts is key to avoid being out-paced by threat actors.

\begin{tcolorbox}[colback=red!5!white, colframe=red!75!black]
  \textbf{Best Practice 2: Triaging.} The most important step in triaging is identifying the relevant file(s) as alerts within the vulnerable file are well concentrated.
\end{tcolorbox}

\begin{table}[ht]
  \centering
  \caption{%
    Per-version CodeQL precision summary. The mean and median (in parentheses) are indicated for each metric.
  }
  \label{tab:locality_versions}

\begin{tabular}{ccccc}
\toprule
\multirow{2}{*}{\shortstack{Version\\ (CVEs)}} & \multicolumn{2}{c}{Hard locality (\%)} & \multicolumn{2}{c}{Soft locality} \\
\cmidrule(lr){2-3} \cmidrule(lr){4-5}
 & Project & File & Project & File \\
\midrule
v2.8.0 (94) & 31.9 (16.7) & 67.9 (75.0) & 62.3 (66.7) & 24.9 (7.5) \\
v2.10.0 (106) & 29.7 (14.3) & 67.0 (70.8) & 58.7 (58.4) & 22.0 (6.2) \\
v2.12.0 (121) & 34.3 (16.7) & 69.8 (100.0) & 62.2 (65.2) & 25.5 (7.4) \\
v2.14.0 (119) & 31.8 (14.3) & 69.5 (100.0) & 60.1 (57.6) & 23.1 (6.2) \\
v2.16.0 (149) & 30.0 (14.3) & 67.7 (85.7) & 60.6 (58.3) & 23.6 (6.2) \\
v2.18.0 (151) & 29.2 (11.1) & 66.1 (75.0) & 60.3 (59.1) & 24.1 (6.9) \\
v2.20.0 (149) & 29.0 (11.1) & 66.6 (75.0) & 59.8 (58.3) & 24.6 (6.7) \\
v2.22.0 (151) & 29.5 (10.3) & 67.2 (80.0) & 60.4 (59.5) & 24.0 (6.7) \\
v2.24.0 (149) & 29.4 (10.0) & 66.9 (75.0) & 60.3 (59.1) & 24.3 (7.1) \\
    \bottomrule
  \end{tabular}
\end{table}

\shortsection{Locality across versions} \autoref{tab:locality_versions} shows the median file and project locality measures. Beyond the raw number of CVEs detected increasing over time, we see that the overall locality only fluctuates lightly across version. CodeQL (like any detector) is subject to the utility-cost trade-off, instantiated here as the compromise between accurate vulnerability detections and small amounts of alerts. It is thus important to characterize the movement of the tool on this trade-off across time due to intended (or accidental) changes, which we will see in the next section.
\subsection{Stability}\label{section:evaluation-stability}
To conclude our evaluation, we ask \ref{rq3}: \textit{is CodeQL's effectiveness stable across versions, and what are the consequences of instability?} Indeed, as explained in \autoref{section:background-codeql}, most software projects using CodeQL actions are bound to use the latest available version alongside the changes it brings, whether they improve or degrade performance (or even create security incidents\footnote{\url{https://nvd.nist.gov/vuln/detail/CVE-2025-24362}}). Following our analysis on the effectiveness and actionability of CodeQL in OSS, we first measure the utility-cost trade-off across versions to characterize short- and long-term directions taken by the tool. Then, we characterize vulnerability detections that were negatively impacted by version changes, representing a blind spot in repositories that use CodeQL without accounting for instability in version changes. We end our analysis by examining how an adversary monitoring these changes can exploit these vulnerabilities and summarize our results into a community best practice on the awareness of such changes.

\begin{figure}
    \centering
    \includegraphics[width=\linewidth]{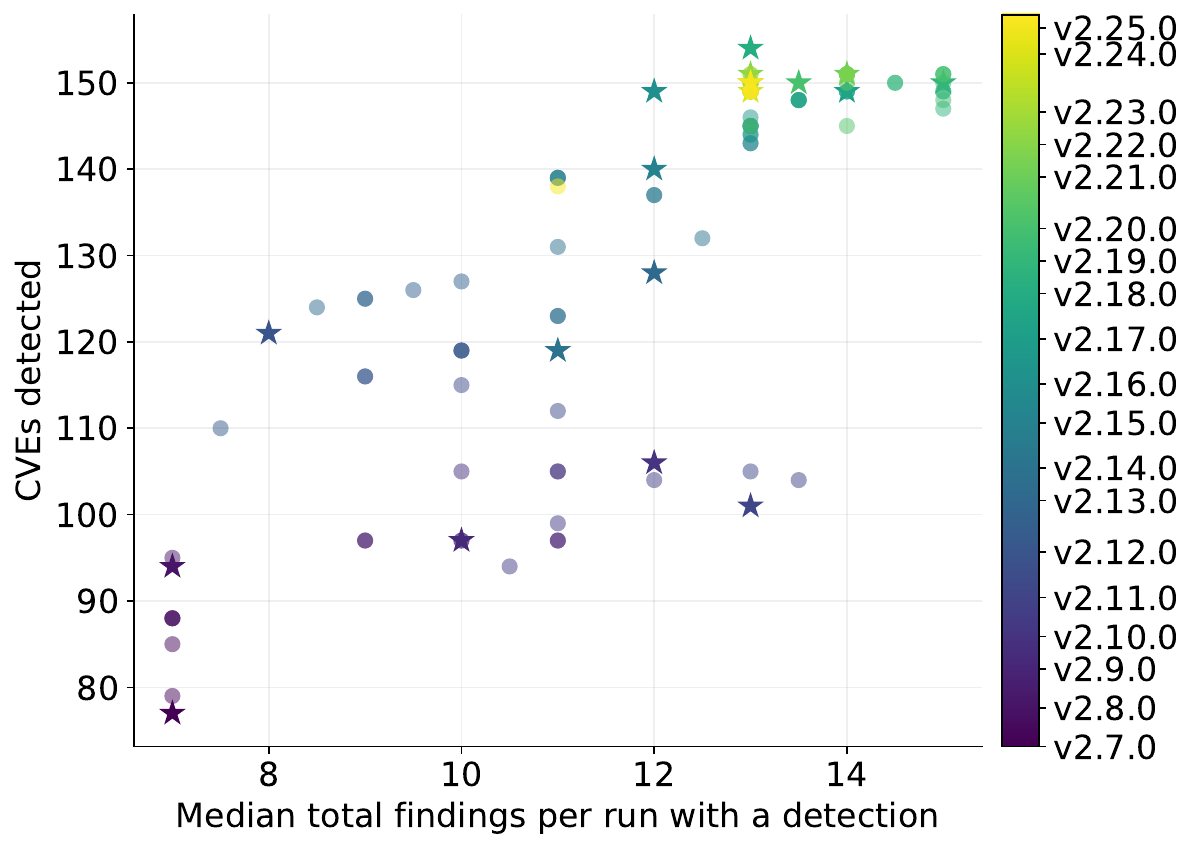}
    \caption{Evolution of CodeQL on the utility-cost trade-off. Star symbols represent minor versions (\texttt{v2.X.0})}
    \label{fig:trade-off-evolution}
\end{figure}

\shortsection{Evolution on the utility-cost trade-off} The crux of the development of a SAST tool like CodeQL---and more generally any classifier or detector---is the position on the utility-cost trade-off: too few detections makes the tool not worth it to adopt while too little precision leads to alert fatigue. Through the evolution of the tool, new queries are added and libraries are tweaked, changing its position on the trade-off space either intentionally (removing a set of queries) or because of low-level changes (\eg in the underlying libraries or the extractors). Through our longitudinal measurement, we can characterize the evolution of CodeQL on this trade-off across versions. \autoref{fig:trade-off-evolution} shows the number of vulnerabilities detected as a function of the median number of alerts, with each dot representing a version. We first note that the first version has both the lowest amount of detections and the lowest median number of alerts. This is consistent with the state of the tool: both the Ruby and the Go ecosystems had issues running this version of the tool. After a phase of high shifts in number of alerts (between v2.8.0 and v2.14.0), the tool converged on a cluster, oscillating between the two ends of the trade-off.

\begin{figure}
    \centering
    \includegraphics[width=\linewidth]{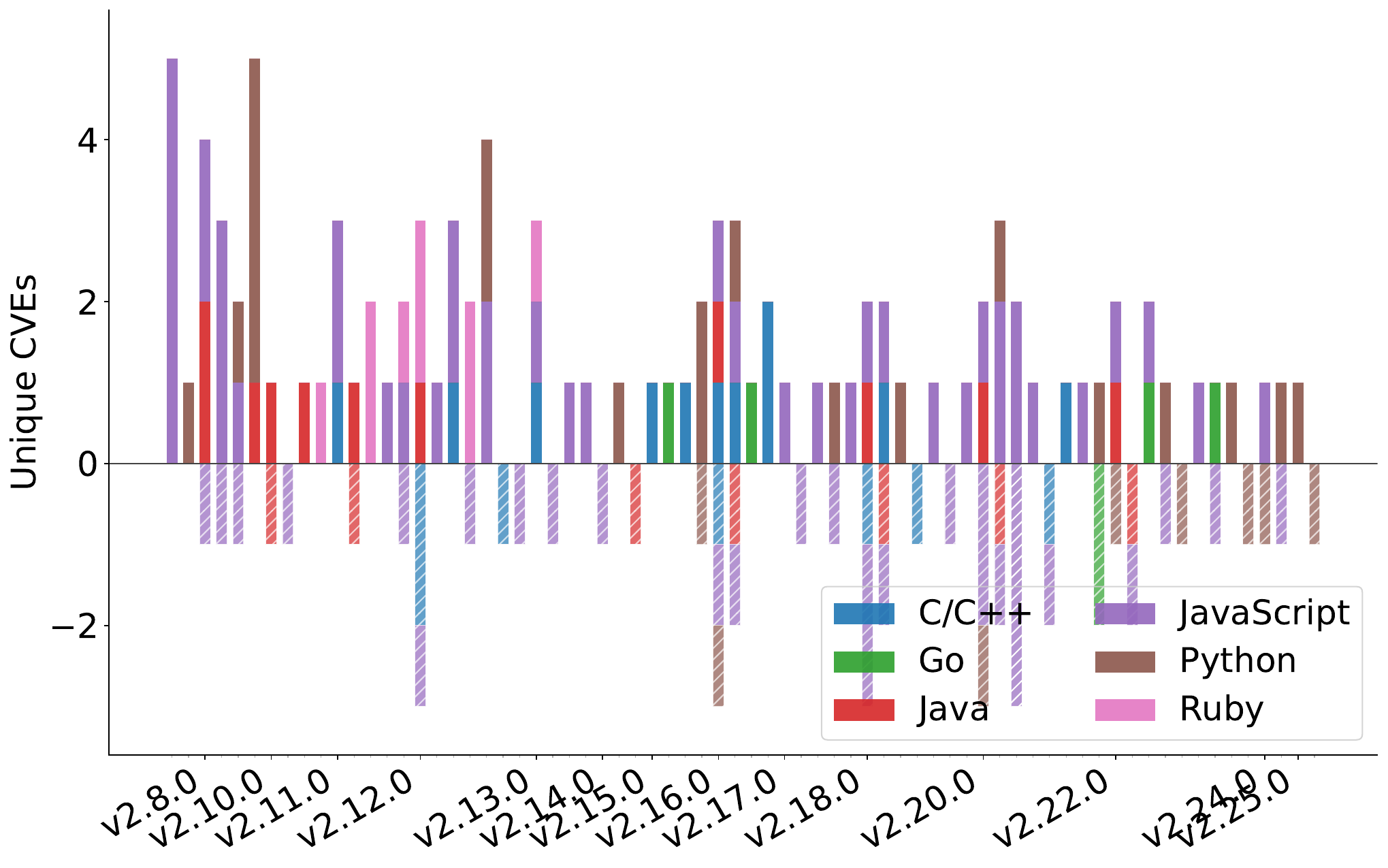}
    \caption{Overall stability of the detections.}
    \label{fig:redetection-drops-lang}
\end{figure}

\shortsection{Detection stability} Given the complexity of CodeQL and its large development (more than 80\;000 observable commits). \autoref{fig:redetection-drops-lang} shows the unique drops (hatched, downward bars) and recoveries of vulnerability detections across versions. Overall, we found that 21 CVEs exhibited such behavior, more than 10\% of the total. Interestingly, we found that many such drops share a common characteristic, they often occur at a new minor version (\eg v2.18.0, v2.20.0, etc.), with the subsequent patch version restoring most detections. We note that some of these minor versions correspond to changes in threat model configurations (v2.18.0) and deprecations of some APIs (v2.20.0). While those particular detection drops might be trivial and quickly fixed, let's recall that CodeQL versions are released every two weeks on average. Thus, a software running the latest version of the CodeQL GitHub Action that introduces a similar vulnerability won't be able to detect it for the same time period. For completeness, we show the timeline of detection for each CVE that was dropped at least once in \autoref{fig:dropped-cves}.

\shortsection{Permanently dropped detections} Interestingly, we observe that a non-negligible amount of CVEs (17) were detected at a given version and then this detection stopped definitely after a change in a version. In particular, CVE-2022-22845 and CVE-2023-27583 were no longer detected starting v2.22.0 for the simple reason that the query that detected them (\texttt{go/hardcoded-credentials}) was removed in v2.21.4, therein dropping the detection for subsequent versions. This version change is part of a larger removal of queries related to secret leakage, a pervasive phenomenon in public GitHub repositories~\cite{meliHowBadCan2019,basakComparativeStudySoftware2023}. For example, the JavaScript equivalent \texttt{js/hardcoded-credentials} was also removed. The main reason for this removal can be attributed to the redundancy of those queries with GitHub Secret Protection~\cite{githubIntroducingGitHubSecret2025}, a different GitHub product that is dedicated to scanning the Git history of a project for hardcoded credentials (\eg API keys, passwords, tokens, etc.) introduced in March 2025---three month before the removal of those queries. While both Code Scanning and Secret Protection are free for public repositories, this removal shows that the sets of queries is not only driven by technical concerns (utility-cost trade-off) but also by product strategy.

\begin{tcolorbox}
    \shortsection{\result{}: Dropped CVEs} Starting v2.22.0 multiple vulnerabilities were no longer detectable through CodeQL as the corresponding queries were removed to leave room for GitHub Secret Protection.
\end{tcolorbox}

\shortsection{Adversaries and stability} As detection stability can oscillate and sometimes create blind spots for the automated pipelines (which use the latest versions), even repositories that use CodeQL might still be vulnerable to threat actors that are aware of the recent changes in the tool and can use them against OSS. It is therefore important for developers to understand the latest changes (or at least those that are pertinent to their project) of the tool that they trust for the security of their software. From this observation, we conclude in the following third and last best practice.

\begin{tcolorbox}[colback=red!5!white, colframe=red!75!black]
  \textbf{Best Practice 3: Awareness and Communication.} SAST tools like CodeQL constantly change. OSS developers need to be aware of these changes and their implications on the software project's security.
\end{tcolorbox}
\section{Discussion}\label{discussion}
\begin{table*}[t!]
\small
\centering
\setlength{\tabcolsep}{4pt}
\renewcommand{\arraystretch}{1.2}
\begin{tabular}{lccccccccccc}
\toprule
\textbf{Study} &
\textbf{Longitudinal} &
\textbf{CVEs} &
\textbf{CWEs} &
\textbf{Vulns} &
\textbf{C/C++} & \textbf{Java} & \textbf{Go} & \textbf{Python} & \textbf{Ruby} & \textbf{JavaScript}  &
\textbf{Dates} \\
\midrule

Lipp et al.~\cite{lippEmpiricalStudyEffectiveness2022}
  & \xmark & \cmark & \cmark & 192
  & \fullcirc & \emptycirc & \emptycirc & \emptycirc & \emptycirc & \emptycirc
  & 2022 \\ \hline

Charoenwet et al.~\cite{charoenwetEmpiricalStudyStatic2024}
  & \xmark & \cmark & \cmark & 319
  & \fullcirc & \emptycirc & \emptycirc & \emptycirc & \emptycirc & \emptycirc
  & 2024 \\ \hline

Shen et al.~\cite{shenFinding709Defects2025}
  & \xmark & \xmark & \xmark & 709
  & \fullcirc & \emptycirc & \emptycirc & \emptycirc & \emptycirc & \emptycirc
  & 2025 \\ \hline

Tracer (Heo et al.)~\cite{kangTRACERSignaturebasedStatic2022}
  & \xmark & \xmark & \xmark & 16 + 5\;388$^*$
  & \fullcirc & \emptycirc & \emptycirc & \emptycirc & \emptycirc & \emptycirc
  & 2022 \\ \hline

Li et al.~\cite{liEvaluatingVulnerabilityDetectability2024}
  & \xmark & \cmark & \cmark & 136 + 30k$^*$
  & \fullcirc & \emptycirc & \emptycirc & \emptycirc & \emptycirc & \emptycirc
  & 2024 \\

\midrule

CASTLE~\cite{dubniczkyCASTLEBenchmarkingDataset2025}
  & \xmark & \xmark & \cmark & 250$^*$
  & \fullcirc & \emptycirc & \emptycirc & \emptycirc & \emptycirc & \emptycirc
  & 2025 \\ \hline

Li et al.~\cite{liComparisonEvaluationStatic2023}
  & \xmark & \cmark & \cmark & 165 + 1\;415+$^*$
  & \emptycirc & \fullcirc & \emptycirc & \emptycirc & \emptycirc & \emptycirc
  & 2023 \\ \hline

Bennett et al.~\cite{bennettSemgrepImprovingLimited2024}
  & \xmark & \cmark & \cmark & 170
  & \emptycirc & \fullcirc & \emptycirc & \emptycirc & \emptycirc & \emptycirc
  & 2024 \\ \hline

Khare et al.~\cite{khareUnderstandingEffectivenessLarge2024}
  & \xmark & \cmark & \cmark & 5\;000$^*$
  & \fullcirc & \fullcirc & \emptycirc & \emptycirc & \emptycirc & \emptycirc
  & 2024 \\ \hline

Zhou et al.~\cite{zhouComparisonStaticApplication2024}
  & \xmark & \cmark & \cmark & 270
  & \fullcirc & \fullcirc & \emptycirc & \fullcirc & \emptycirc & \emptycirc
  & 2024 \\ \hline

Brito et al.~\cite{britoStudyJavaScriptStatic2023}
  & \xmark & \cmark & \cmark & 957
  & \emptycirc & \emptycirc & \emptycirc & \emptycirc & \emptycirc & \fullcirc
  & 2023 \\ 

\midrule

\textbf{Ours}
  & \cmark & \cmark & \cmark & 3\;993
  & \fullcirc & \fullcirc & \fullcirc & \fullcirc & \fullcirc & \fullcirc
  & 2019--2025 \\
\bottomrule
\end{tabular}
\caption{Comparison of empirical CodeQL / SAST tool evaluation studies. For each study, we analyze each axis of the evaluation: the use of real-world vulnerabilities (CVEs), the characterization of weaknesses (CWEs), the amount of samples, the scope of languages, and the time range studied.
$^*$~includes synthetic samples.}
\label{table:prior-work-comparison}
\end{table*}

\subsection{Codependency and OSS}
The key aspect of this study is the study of the evolution of the SAST tool. However, the same way that the tool evolves, OSS ecosystems---which depend on it---change over time as well. Projects such as CodeQL require a lot of effort to ensure that it is precise and helpful for developers. This effort is continuous since all software ecosystems change while others are introduced: a new language or an update to an existing one need to be reflected onto CodeQL so that projects can use it. Similarly, the knowledge base on vulnerabilities in code and their patterns is in constant expansion, therein implicitly shaping the development of static analysis tools to stay on par with developers' expectations.

\shortsection{Ecosystem representativeness} Due to this interdependency between OSS and SAST, an implicit supply and demand model happens: adoption rates of the tool across language ecosystems vary significantly (see \autoref{fig:codeql-adoption}) due to improvements of the tool. In particular, the language ecosystem that historically adopted CodeQL the most is Go with more 30\% of repositories (in a stratum of repositories with more than 400 stars and with at least one commit after January 2025). In this work, we focused on one longitudinal axis and ignored the longitudinal axis of OSS evolution. An extension to this work would be to consider both the SAST and the OSS as two codependent entities that follow this supply and demand model.

\subsection{Developing a SAST}

\shortsection{Language expertise} Static analysis requires a significant expertise in the target language, as it can range from simple pattern matching (hardcoded credentials) to deep taint tracking (SQL injection), not only demanding a fine understanding of the language but also an awareness of the language's ecosystem (so that sources, sinks, and sanitizers are up-to-date). Therefore, a static analysis tool like CodeQL that spans many languages is not realizable by a single person, as shown by the more than 300 contributors to the CodeQL repository. We found that well established language ecosystems (Java, C/C++, Python) exhibit the highest number of contributors while the most recently supported languages like Ruby and Go are significantly smaller (\eg 24 for Go against 80 for Java, see \autoref{appendix:upset-contributor} for more details). This highlights a mismatch between development and adoption (see \autoref{section:background-codeql}) across languages.

\shortsection{AI and SAST} Recent improvements in artificial intelligence have led to a flourishing in combinations between CodeQL and LLMs~\cite{wangQLCoderQuerySynthesizer2026,liIRISLLMAssistedStatic2025}. Throughout the development of the tool, the trade-off between accuracy and precision can materialize in explicit distinction like the default and the extended suites or in subtle changes in the queries. Using LLMs that leverage the program semantic extraction and analysis of CodeQL is thus a great opportunity to have better queries, but also to navigate this trade-off space on a per-project basis as different projects might have different needs (\eg a single-maintainer project might prefer few very precise queries compared to a large project that suffered from many CVEs).

\section{Related Work}\label{section:related-work}
Static analysis tools have undergone multiple evaluations either as part of a global assessment of the tools or to compare to new approaches. We summarize in \autoref{table:prior-work-comparison} prior works that evaluate CodeQL---among other tools---alongside the dataset used and languages considered and detail below key differentiators of our work.

\shortsection{Longitudinality} In all evaluations, static analysis tools are treated as immutable, fixed programs. However, tools such as CodeQL evolve: new versions improve queries logic, adds more ecosystem support, and fix bugs. While prior work gives an accurate measurement of the performance of a SAST tool, the findings, while true at the time, might become stale as the tool improves. This can lead to confusion as the version number is often not specified~\cite{shenFinding709Defects2025,bennettSemgrepImprovingLimited2024,khareUnderstandingEffectivenessLarge2024}.

\shortsection{Classification vs.\ Usefulness} Vulnerability detection has often been framed as a binary classification 
problem~\cite{ghaffarianSoftwareVulnerabilityAnalysis2017}, even more so with the increasing adoption of machine learning 
approaches~\cite{yangDLAPDeepLearning2025}. Under this framing, a tool is evaluated on whether it detects a vulnerability or not,  without considering whether the generated alerts are actionable for a developer. However, a SAST tool's goal is to inform developers of potential security weaknesses in a way they can act on---a detection that is buried among hundreds of unrelated alerts in the wrong region of the codebase offers limited practical value, regardless of its binary correctness. Prior works that measure precision and recall at the vulnerability level~\cite{lippEmpiricalStudyEffectiveness2022, 
liComparisonEvaluationStatic2023, 
khareUnderstandingEffectivenessLarge2024} do not capture this distinction. Our locality metrics address this gap by measuring where alerts appear relative to the vulnerable code, rather than whether a detection occurs at all.

\shortsection{Evaluation Scope} Prior works have mostly considered Java~\cite{wangQLCoderQuerySynthesizer2026,huQLProAutomatedCode2025,liComparisonEvaluationStatic2023}, C (and C++)~\cite{lippEmpiricalStudyEffectiveness2022,charoenwetEmpiricalStudyStatic2024,shenFinding709Defects2025} or both~\cite{khareUnderstandingEffectivenessLarge2024} as the languages of choice for such evaluations. This follows the large ecosystem of static analysis with well established benchmarks~\cite{owaspfoundationOWASPBenchmark,bolandJuliet11Java2012}. On the other hand, highly popular languages such as JavaScript, Python, or Go have remained underrepresented. In contrast, there hasn't been a strong consensus on what dataset to use. Synthetic benchmarks such as the Juliet Test 
Suite~\cite{bolandJuliet11Java2012}, the OWASP Benchmark~\cite{owaspfoundationOWASPBenchmark}, or CASTLE~\cite{dubniczkyCASTLEBenchmarkingDataset2025} offer controlled comparison points and are used by several works~\cite{khareUnderstandingEffectivenessLarge2024,liComparisonEvaluationStatic2023,dubniczkyCASTLEBenchmarkingDataset2025}, but their synthetic nature limits generalizability to real-world 
OSS, and they are often restricted to one or two languages. Li et al.~\cite{liEvaluatingVulnerabilityDetectability2024} combine synthetic and real-world samples for C/C++, providing broader coverage but still within a single language. Works grounded in real-world CVEs~\cite{lippEmpiricalStudyEffectiveness2022, 
kangTRACERSignaturebasedStatic2022, britoStudyJavaScriptStatic2023} offer higher ecological validity but are typically constrained to one or two languages. Our study uses CVEfixes to span six languages with real-world vulnerabilities, combining the validity of CVE-based evaluation with the breadth of multiple languages.
\section{Conclusion}\label{conclusion}
In this work, we performed the largest academic longitudinal analysis on a static analysis tool to date. We found that \NumDetectedCVEsInTime CVEs could have been identified using the CodeQL version available at the time, prompting open-source software to make use of available static analysis tools. Through the prism of project- and file-level locality, our results show that the tool can be seen as accurate provided that developers know what part of the codebase should be focused on. Finally, our analysis of the stability of the tool revealed that there can be drops of detections with new releases, emphasizing that developers should be aware of the tool's evolution.

\begin{acks}

\ifanonymous
Anonymized for review.
\else

\textbf{Funding acknowledgment:} This material is based upon work supported by the National Science Foundation under Grant No. CNS-2343611 and by PRISM, one of seven centers in JUMP 2.0, a Semiconductor Research Corporation (SRC) program sponsored by DARPA. Any opinions, findings, and conclusions or recommendations expressed in this material are those of the author(s) and do not necessarily reflect the views of the National Science Foundation.

\fi
\end{acks}

\bibliographystyle{ACM-Reference-Format}
\bibliography{refs}

\appendix 

\section{Appendix}

\subsection{Contributors and Language expertise}\label{appendix:upset-contributor}
\begin{figure}[h!]
    \centering
    \includegraphics[width=.8\linewidth]{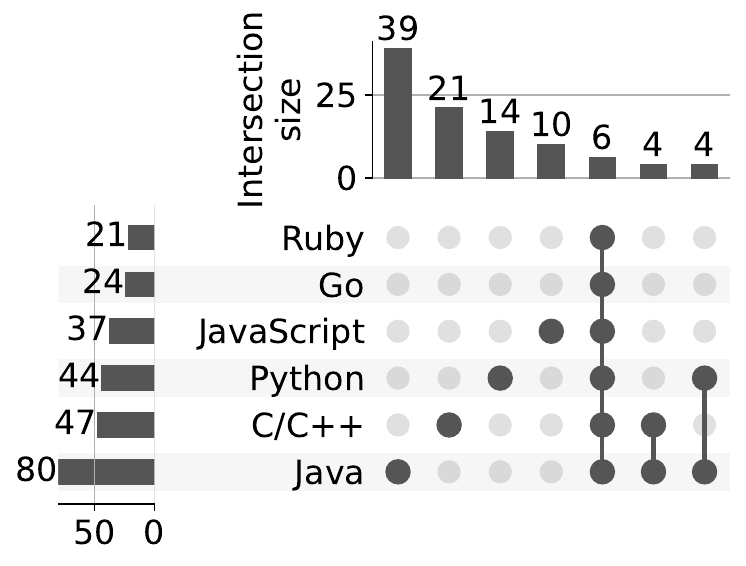}
    \caption{UpSet plot of the contributors per language, truncated at 10 intersections for readability.}
    \label{fig:language-expertise}
\end{figure}

\shortsection{Intersection of expertise} \autoref{fig:language-expertise} shows an UpSet~\cite{lexUpSetVisualizationIntersecting2014} plot of developers who contributed to security queries or language-specific libraries for each language. The intersection size for each language expertise intersection (as represented by the linked dots) is shown on the top histogram, while the histogram on the left shows the total number of unique developers for a given language (regardless of the focus or spread across other languages). Notably, a core of 6 contributors spans all languages.

\newpage
\subsection{Dropped CVEs}
\begin{figure}[h!]
    \centering
    \includegraphics[width=.9\columnwidth]{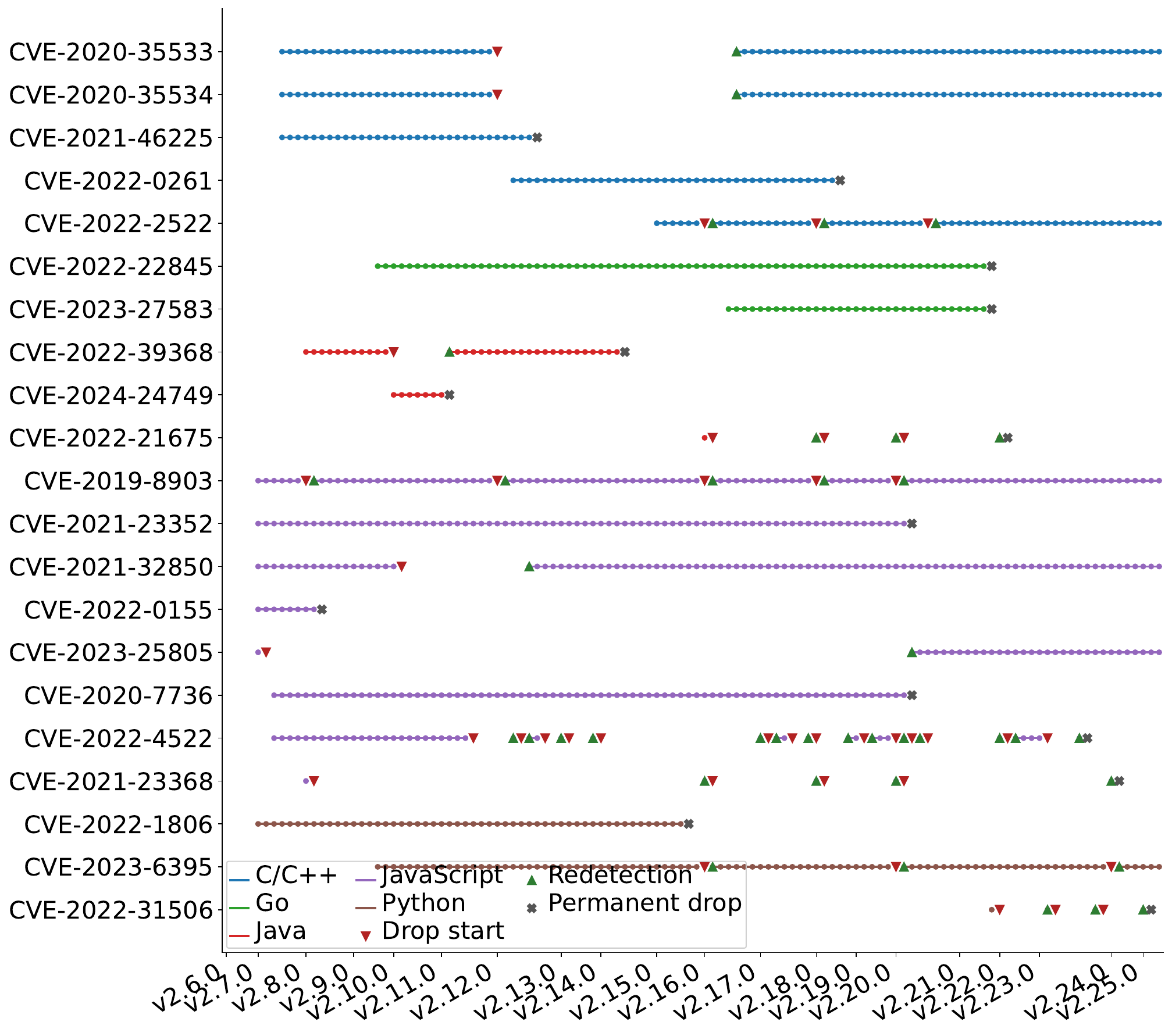}
    \caption{Vulnerability detections stability across releases}
    \label{fig:dropped-cves}
\end{figure}

\autoref{fig:dropped-cves} shows the timeline of vulnerability detections and drops across version for vulnerability with at least one drop.

\end{document}